\documentclass[preprint,prd,nofootinbib,tightenlines,amsmath,amssymb]{revtex4}
\usepackage[dvipdfmx]{graphicx}
\usepackage{bm}
\usepackage{amsmath}
\usepackage{hhline}
\usepackage{color}
\usepackage{xcolor}
\usepackage{slashed}
\usepackage{relsize}

\newcommand{\Eq}{&=&}

\newcommand{\white}[1]{{\color[rgb]{1,1,1} #1}}
\newcommand{\black}[1]{{\color[rgb]{0,0,0} #1}}

\newcommand{\hs}[1]{{\hspace{#1}}}
\newcommand{\vs}[1]{{\vspace{#1}}}

\begin{document}

\vspace*{3em}

\preprint{KIAS-P21006}%

\title{Explaining the MiniBooNE anomalous excess \\[0.1cm] via leptophilic ALP-sterile neutrino coupling} 

\author{
Chia-Hung Vincent Chang,$^{1}$\footnote[1]{chchang@phy03.phy.ntnu.edu.tw}
Chuan-Ren Chen,$^{1}$\footnote[2]{crchen@ntnu.edu.tw}
Shu-Yu Ho,$^{2}$\footnote[3]{phyhunter@kias.re.kr}
and Shih-Yen Tseng$^{\,3}$\footnote[4]{shihyen@hep-th.phys.s.u-tokyo.ac.jp}
}

\affiliation{
${}^{1}$Department of Physics, National Taiwan Normal University, Taipei 116, Taiwan \vspace{3pt} \\
${}^{2}$Korea Institute for Advanced Study, Seoul 02455, Republic of Korea \vspace{3pt} \\
${}^{3}$Department of Physics, Faculty of Science, The University of Tokyo, Bunkyo-ku, Tokyo 113-0033, Japan
\vspace{3ex}}

\begin{abstract}
\vs{0.2cm}
Recently, the MiniBooNE experiment at Fermilab has updated the results with increased data and reported an excess of $560.6 \pm 119.6$ electronlike events ($4.7\sigma$) in the neutrino operation mode.\\In this paper, we propose a scenario to account for the excess where a Dirac-type sterile neutrino, produced by a charged kaon decay through the neutrino mixing, decays into a leptophilic axionlike particle ($\ell$ALP) and a muon neutrino.\,\,The electron-positron pairs produced from the $\ell$ALP decays can be interpreted as electronlike events provided that their opening angle is sufficiently small. In our framework, we consider the $\ell$ALP with a mass $m^{}_a = 20\,\text{MeV}$ and an inverse decay constant $c^{}_e/f^{}_a = 10^{-2}\,\text{GeV}^{-1}$, allowed by the astrophysical and experimental constraints.\,\,Then, after integrating the predicted angular or visible energy spectra of the $\ell$ALP to obtain the total excess event number, we find that our scenario with sterile neutrino masses  within $150\,\text{MeV}\lesssim m^{}_N \lesssim 380 \,\text{MeV}$ ($150\,\text{MeV}\lesssim m^{}_N \lesssim 180 \,\text{MeV}$) and neutrino mixing parameters between $10^{-10} \lesssim |U_{\mu 4}|^2 \lesssim 10^{-8}$ ($3\times 10^{-7} \lesssim |U_{\mu 4}|^2 \lesssim 8 \times10^{-7}$) can explain the MiniBooNE data.
\end{abstract}

\maketitle

\section{Introduction}\label{sec:1}

Since the groundbreaking discovery of the neutrino oscillations at the Super-Kamiokande experiment in 1998\,\cite{Fukuda:1998mi}, numerous measurements have provided clear evidence that neutrinos have nonzero masses and the mass eigenstates are an admixture of the flavor eigenstates\,\cite{Hernandez:2017txl}. Even though the mass generation mechanism and the mass ordering are still unknown,\,\,it is well understood that the assumption that neutrinos are of three different flavors ($\nu_e,\nu_\mu,\nu_\tau$) with two mass splittings  and three mixing angles gives a good fit to most of the neutrino data, including solar neutrinos, atmospheric neutrinos, long-baseline, and reactor experiments\,\cite{Hernandez:2017txl}.  

On the other hand, there are some long-standing anomalies which suggest the existence of nonstandard neutrinos.\,\,For instance, an excess of $\bar{\nu}_\mu\to\bar{\nu}_e$ appearance observed by the short-baseline experiment LSND Collaboration \cite{Aguilar:2001ty} indicates the presence of a fourth-flavor neutrino, most likely a sterile neutrino $\nu_s$, participating in the neutrino oscillation scenario with a much larger mass splitting of eV scale.\,\,With the similar design $L_\nu/E_\nu \sim 1\,\text{m}/\text{MeV}$, where $L_\nu$ and $E_\nu$ are the travel distance and energy of neutrino, respectively, MiniBooNE at FermiLab is built up to confirm or disprove the anomaly reported by LSND. Based on the $\nu_e$ and $\bar{\nu}_e$ appearance data collected from 2002 to 2019, MiniBooNE reports excesses of 561 events in neutrino mode and 77 events in antineutrino mode, which corresponds to $4.8{}^{}\sigma$ effect in total~\cite{MiniBooNE:2020pnu}.\,\,Combining with the LSND result, the significance even reaches $6.1{}^{}\sigma$. Assuming one sterile neutrino and applying a two-neutrino oscillation model, MiniBooNE reports the best-fit point for data with the mass splitting $\Delta m^2 = 0.043\,\rm{eV^2}\hs{-0.05cm}$ and the mixing angle $\sin^2 2\theta = 0.807$~\cite{MiniBooNE:2020pnu}, requiring $\Delta m^2 \gtrsim 0.03\,\rm{eV^2}$ at $90\%$ C.L., in agreement with LSND. However, the introduction of an eV-scale sterile neutrino, while generating the $\nu_{\mu} \rightarrow \nu_{e}$ appearance, also gives rise to the $\nu_e$ disappearance at the short-baseline experiment, which unfortunately is not observed.\,\,That is to say the parameter region in the sterile neutrino scenario favored by the MiniBooNE result is incomparable with the global fit for the other neutrino data~\cite{Dentler:2018sju,Boser:2019rta,Diaz:2019fwt}.  

Therefore, there is a significant interest in alternative explanations of the excess~\cite{Bertuzzo:2018itn,Fischer:2019fbw,Dentler:2019dhz}. Intriguingly, the MiniBooNE detector is unable to distinguish the single electron signal of a $\nu_e$ charged-current quasielastic scattering ($\nu_e {}^{} n \to p {}^{}{}^{} e^-$) from photons or a collimated $e^+e^-$ pair. Several efforts have been devoted to the possibility of light exotic particles that decay inside the detector into photons or $e^+e^-$ pairs to camouflage the electron signals.\,\,For example, a plausible alternative is a decaying sterile neutrino that is produced in charged meson decays.\,\,Because of the mixings with the active neutrinos, the sterile neutrino can be produced, if kinematically allowed, in the decay of mesons when the proton beam hits the target.\,\,The sterile neutrino could decay into standard model (SM) particles, e.g. $\nu_s\to\nu_\beta \gamma$ with $\beta = e,\mu$, and $\tau$~\cite{Pal:1981rm,Barger:1995ty}.\,\,However, its lifetime is usually long enough such that $\nu_s$ can be regarded as a stable particle in the short-baseline experiments.\,\,If new interactions are introduced, the sterile neutrino would have more decay modes and could decay within the length scale of the MiniBooNE experiment, even decay promptly.\,\,Then, the decay of sterile neutrino into photons or $e^+e^-$ pairs inside the detector could possibly provide the excess reported by MiniBooNE~\cite{Fischer:2019fbw}.   

However, it has been pointed out in Ref.\,\cite{Jordan:2018qiy} that it is difficult for them to fit both the angular and energy distributions of the excess events.\,\,The key obstacle is that if the light new particle decays visibly the total momentum of the $\nu_e$-like products will be equal to that of the light new particle.\,\,For this new particle to enter the MiniBooNE detector, the track angle must be small, and thus the angular spectrum of the excess events is forward peaked.\\Nevertheless, the MiniBooNE data have significant excess even for $\cos \theta_e < 0.8$.\,\,This tension can be alleviated if the new particle decays semivisibly since the invisible product could take away some transverse momentum.\,\,Following this strategy, Ref.\,\cite{Fischer:2019fbw} proposes a scenario where the sterile neutrino decays into a photon and a light neutrino, $\nu_s\to\nu_\beta \gamma$.\,\,The angular distribution is still more forward peaked compared to data.\,\,It was also proposed in Refs.\,\cite{Bertuzzo:2018itn,Ballett:2018ynz} that the $\nu_{\mu}$ may scatter with nucleons inside the detector via new physics to produce a sterile neutrino, which subsequently decays into $e^+e^-$ pairs, mimicking excess events.\,\,The scenario seems to have a less forward peaked angular distribution of the excesses. 

\begin{figure}[t]
\includegraphics[scale=0.38]{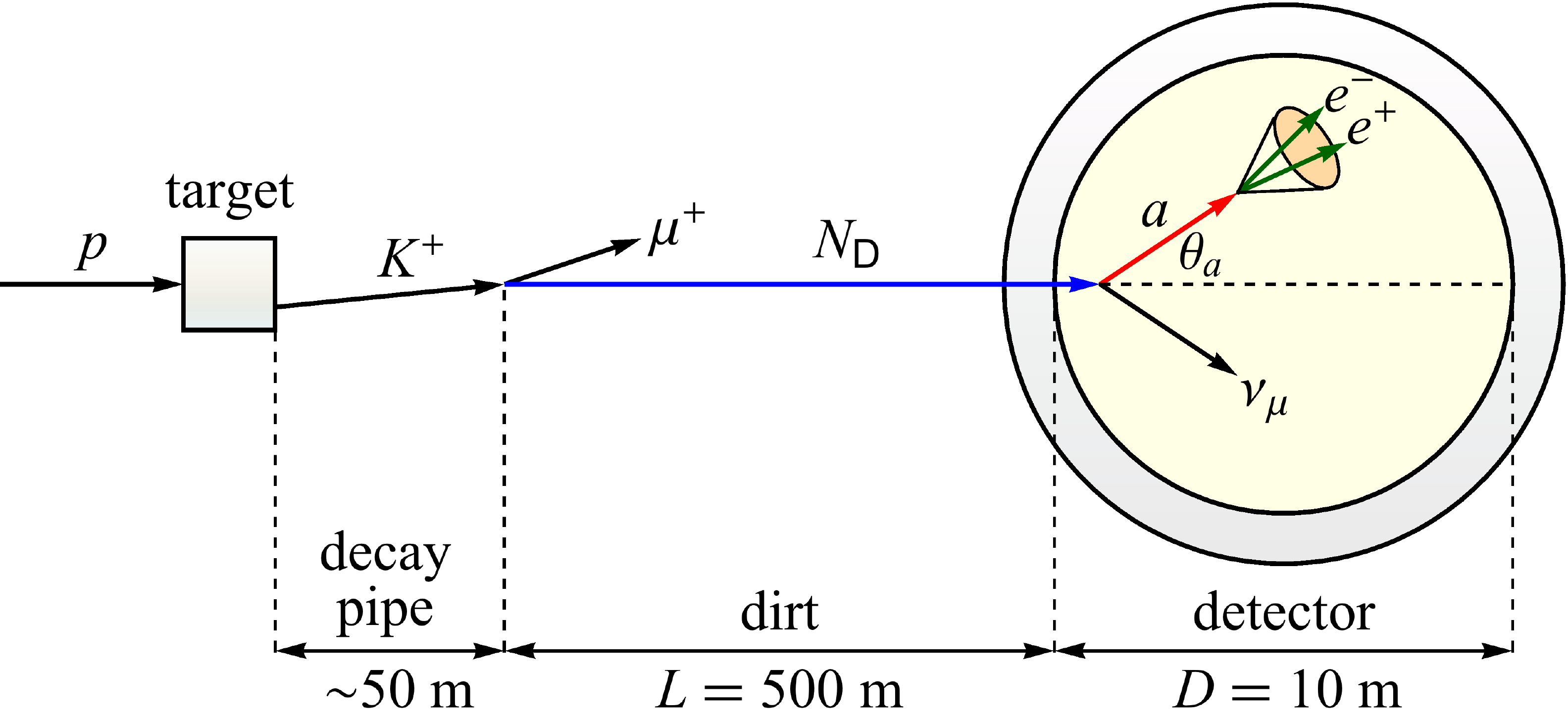}  
\caption{The illustration of our setup to explain the MiniBooNE excess electronlike events in the $\ell$ALP model, where $L$ is the travel distance of the sterile neutrino produced by the charged kaon decays, $D$ is the diameter of the MiniNooBE detector, and $\theta_a$ is the scattering angle of the $\ell$ALP produced from the sterile neutrino decay.\,\,If the angular aperture of  electron-positron pairs produced from the $\ell$ALP decays is sufficiently small, they can be treated as electronlike events.}
\label{fig:setup}
\end{figure}

In this work, we tend to explain the MiniBooNE excess by a sterile neutrino $N_\textsf{D}$ of mass around 100$\,\sim\,$400 MeV and a ${\cal O}(10)\,\text{MeV}$ leptophilic axionlike particle ($\ell$ALP), $a$~\cite{Han:2020dwo}.\,\,The sterile neutrino is produced in the decay of kaon from the target via its mixing with the $\nu_\mu$.\,\,Then, it travels about $500\,\text{m}$ and decays semivisibly into a muon neutrino and a $\ell$ALP, which in turn decays into an electron-positron pair in the detector, as sketched in Fig.\,\ref{fig:setup}. Our calculation shows that it is possible to obtain a rather mild forward peaked angular distribution of excess.\,\,In general, the mass of an axionlike particle and its couplings to the SM fields are strictly constrained by beam-dump experiments, astrophysical observations, and rare decays of mesons.\,\,However, most of the productions of axionlike particles in the aforementioned experiments rely on the couplings to the SM quarks.\,\,Since we consider a $\ell$ALP that interacts with the SM leptons only, as a result, the relevant bounds are placed by supernova 1987A, electron beam-dump experiment E137, and electron $(g-2)^{}_e$ anomaly.\,We will discuss these constraints later.

The structure of this paper is organized as follows.\,\,In the next section, we introduce the effective Lagrangian of the $\ell$ALP, focusing on the couplings to electrons and photons.\,\,We also discuss the decay modes of sterile neutrino and $\ell$ALP. Section III is the discussion about the constraints of parameters in our model, including the supernova 1987A, E137, electron magnetic dipole moment anomaly, and rare kaon decay.\,\,In Sec.\,\ref{sec:4}, we demonstrate how we estimate the excess of $\nu_e$-like events and show our fits to the MiniBooNE results.\,\,The last section is devoted to discussion and conclusions.

\section{Theoretical Framework}\label{sec:2}
\subsection{Sterile neutrino and leptophilic ALP}\label{subsec:1}

In our setup, we add one Dirac-type sterile neutrino, $\nu^{}_{\textsf D}$, to the SM neutrino sector.\,\,As usual, the neutrino flavor eigenstates could be transformed into the mass eigenstates by a unitary matrix ${}^{}U$.\,\,Explicitly, one can express the neutrino flavor eigenstate $\nu^{}_\beta$ as a superposition of the neutrino mass eigenstates $\nu^{}_{jL},N^{}_\textsf{D}$\,\cite{Bertuzzo:2018itn},
\begin{eqnarray}\label{numix}
\nu^{}_\beta
\,=\, 
\sum_{j{}^{}={}^{}1}^{3}  U^{}_{\beta j} {}^{}\nu^{}_{jL} + U^{}_{\beta 4} N^{}_\textsf{D}
~,
\end{eqnarray}
where $\beta \,=\, e,\mu,\tau,\textsf{D}$ and  $j \,=\, 1,2,3 $ are the flavor and generation indices, respectively.

To account for the MiniBooNE excess, we also introduce a $\ell$ALP, $a$, which only couples to the leptons but not quarks.\,\,What is relevant for us is the interactions of $\ell$ALP with the sterile neutrino and electron.\,\,Assuming, for simplicity, the interaction is diagonal in the flavor eigenstates of leptons, the effective Lagrangian density can be written as~\cite{Alves:2019xpc}
\begin{eqnarray}\label{Lint}
{\cal L}^{}_{a \ell} 
\,=\, 
-\,\frac{\partial_\mu a}{2f^{}_a} 
\Big(
c^{}_{N} {}^{} \overline{\nu^{}_\textsf{D}} {}^{} \gamma^\mu \gamma^5 \white{\overline{\black{\nu^{}_\textsf{D}}}} 
+
c^{}_{e} {}^{} \overline{e} {}^{} \gamma^\mu \gamma^5 e 
\Big)~,
\end{eqnarray}
where $f^{}_a$ is the $\ell$ALP decay constant and $c^{}_{N}$ and $c^{}_e$ are dimensionless parameters of order of unity. 
Notice that the diagonal $\ell$ALP-vector current interactions give no physical effect due to the conservation of the vector currents; thereby, we omit $\partial_\mu a \, \bar{\ell}\gamma^\mu \ell$ interactions.

Plugging Eq.\,\eqref{numix} into Eq.\,\eqref{Lint}, we then obtain a mixing between the mass eigenstates of the sterile and active neutrinos with the $\ell$ALP coupling
\begin{eqnarray}\label{aNnu}
{\cal L}^{}_{a\ell} 
\,\supset\, 
- {}^{}{}^{} c^{}_{N} \frac{\partial_\mu a}{2f^{}_a} 
\Big( 
U^{}_{\textsf{D}j}U^{\ast}_{\textsf{D}4} \overline{N^{}_\textsf{D}} {}^{} \gamma^\mu\gamma^5 P^{}_L \nu^{}_{j} 
+
\text{H.c.}
\Big) ~,
\end{eqnarray}
where $P^{}_L = \frac{1}{2}(1-\gamma^5)$ is the left-hand project operator.\,\,This term will be responsible for the decay of the sterile neutrino into the $\ell$ALP inside the detector.

Besides the above couplings, the $\ell$ALP can also interact with photons via the one-loop triangle diagrams and chiral anomaly.\,\,We can rewrite the $\ell$ALP-electron coupling in Eq.\,\eqref{Lint} by applying the anomaly equation for the divergence of the axial-vector current
\begin{eqnarray}\label{agamma}
c^{}_{e} \frac{\partial_\mu a}{2f^{}_a} \,\overline{e} {}^{} \gamma^\mu \gamma^5 e
\,=\,
- \, c^{}_e \frac{m^{}_e}{f^{}_a} {}^{}{}^{} a {}^{}{}^{} \overline{e} {}^{}{}^{}i{}^{} \gamma^5 e
+ 
c^{}_e \frac{\alpha}{4\pi} \frac{a}{f^{}_a} F_{\mu\nu} \widetilde{F}^{\mu\nu} ~,
\end{eqnarray}
where $\alpha {}^{}\simeq {}^{} 1/137$ is the fine structure constant, $F_{\mu\nu} = \partial_\mu A_\nu - \partial_\nu A_\mu$ is the field strength tensor of photon, and $\widetilde{F}^{\mu\nu} = \frac{1}{2}\epsilon^{\mu\nu\rho\sigma} F_{\rho\sigma}$ is its dual tensor with $\epsilon^{0123} = +1$.\,\,From Eq.\,\eqref{agamma}, the effective interaction between the $\ell$ALP and photons equals~\cite{Bauer:2017ris}
\begin{eqnarray}\label{Lag}
{\cal L}_{a\gamma} 
\,=\, 
- {}^{}{}^{} \frac{1}{4} g^{}_{a\gamma\gamma} {}^{} a {}^{} F_{\mu\nu} \widetilde{F}^{\mu\nu} 
\end{eqnarray}
with $g^{}_{a\gamma\gamma}$ the $\ell$ALP-photon coupling of the form 
\begin{eqnarray}\label{gagg}
g^{}_{a\gamma\gamma}  \,=\, \frac{\alpha}{\pi}\frac{c^{}_e}{f^{}_a}
\bigg|1-{}^{}{\cal F}\bigg(\frac{m_a^2}{4 m_e^2}\bigg)\bigg|
~,
\end{eqnarray}
where $m^{}_a$ is the $\ell$ALP mass, the factor of $1$ in the absolute value comes from the anomaly term, and ${\cal F}(z)$ is the loop function whose form depends on the argument.\,\,To explain the MiniBooNE excess, we will assume that $m^{}_a > 2 m^{}_e$.\,\,In this case, the loop function reads~\cite{Calibbi:2020jvd}
\begin{eqnarray}\label{Ftau}
{\cal F}(z >1) \,=\, \frac{1}{z} {}^{} {\arctan}^2
\scalebox{1.1}{\bigg(}
\frac{1}{\sqrt{1/z-1}}
\scalebox{1.1}{\bigg)}
~.
\end{eqnarray}
Note that we have $g^{}_{a\gamma\gamma} \,\simeq\, 2.32 \times 10^{-3} (c^{}_e/f^{}_a)$ for $m^{}_a \gg m^{}_e{}^{}$, since ${\cal F}(z \gg 1) \to 0{}^{}$.\,\,We will use this interaction to calculate the photophilic decay of the $\ell$ALP in the next subsection.

\subsection{Decay width}\label{subsec:2}

We propose that at MiniBooNE, when the proton beam hits the target, the charged $K$ meson produced decays into a sterile neutrino through its mixing with the muon neutrino in Eq.\,\eqref{numix}.\,\,In our study, we assume that the sterile neutrino is much heavier than the $\ell$ALP. Thus, the sterile neutrino can decay into a $\ell$ALP and a light neutrino as $N^{}_{\textsf D} \to a + \nu^{}_{jL}$. Using Eq.\,\eqref{aNnu}, the decay rate of the sterile neutrino into $a + \nu's$ is calculated as 
\begin{eqnarray}\label{decayN}
\Gamma^{}_{N^{}_\textsf{D} \to a {}^{} \nu} 
\Eq
\Gamma^{}_{\bar{N}^{}_\textsf{D} \to  a {}^{} {\bar\nu}} 
\,=\, 
\sum_{j{}^{}={}^{}1}^{3} \Gamma^{}_{N^{}_\textsf{D} \to  a {}^{} \nu^{}_{jL}}
\,=\,
\frac{c^2_N  |U_{\mu{\rm 4}}|^2 m^3_N}{128\pi f_a^2}\bigg(1-\frac{m^2_a}{m^2_N}\bigg)^{\hs{-0.13cm}2}
\nonumber\\[0.1cm]
& \simeq &
1.36\times 10^{-15} \,\text{MeV}
\bigg(\frac{|U_{\mu{\rm 4}}|}{1\times 10^{-5}}\bigg)^{\hs{-0.15cm}2}
\bigg(\frac{m^{}_N}{380\,\text{MeV}}\bigg)^{\hs{-0.15cm}3}
\bigg(\frac{f^{}_a}{100\,\text{GeV}}\bigg)^{\hs{-0.15cm}-2}
~,
\end{eqnarray}
where $m^{}_N$ is the sterile neutrino mass.\,\,In Eq.\,\eqref{decayN}, we have used the unitary condition and symmetry property of ${}^{}U$ and also assumed, for simplicity, $|U_{e{\rm 4}}|^2,|U_{\tau{\rm 4}}|^2 \ll |U_{\mu{\rm 4}}|^2 \ll 1$.\,With the mixing parameter $U_{\mu{\rm 4}}$, the sterile neutrino can also decay into a muon plus a charged pion or a muon neutrino plus a neutral pion\,\,:\,\,$N^{}_{\textsf D} \to \mu^\pm \pi^\mp$ or $\nu^{}_\mu \pi^0$ if it is kinematically allowed.\,\,The corresponding decay rates have been estimated in Ref.\,\cite{Bondarenko:2018ptm} as
\begin{eqnarray}
\Gamma^{}_{N^{}_\textsf{D}\to \mu (\nu_\mu) \pi} 
\Eq
\frac{G^2_\textsf{F} f^2_\pi |U_{\mu{\rm 4}}|^2 m^3_N}{32\pi} {}^{} {\cal K}\big[m^{}_\pi, m^{}_\mu(0), m^{}_N\big]
\nonumber\\[0.1cm]
&\simeq&
1.25\times 10^{-22} \,\text{MeV}
\bigg(\frac{|U_{\mu{\rm 4}}|}{1\times 10^{-5}}\bigg)^{\hs{-0.15cm}2}
\bigg(\frac{m^{}_N}{380\,\text{MeV}}\bigg)^{\hs{-0.15cm}3}
~,
\end{eqnarray}
where $G^{}_\textsf{F} \simeq 1.166 \times 10^{-5}\,\text{GeV}^{-2}$ is the Fermi coupling constant, $f^{}_\pi \simeq 130\,\text{MeV}$ is the pion decay constant, and ${\cal K}$ is an ${\cal O}(1)$ dimensionless kinematical function~\cite{Bondarenko:2018ptm}.\,\,Apparently, these decay channels are subdominant in comparison with $N^{}_{\textsf D} \to a \nu$ unless $f^{}_a \gtrsim 300\,\text{TeV}$. Hence, the dominant decay mode of the sterile neutrino after it arrives at the detector is a $\ell$ALP plus a light neutrino.

\begin{figure}[t!]
\vs{-0.3cm}
\hs{-1.0cm}
\centering
\includegraphics[scale=0.5]{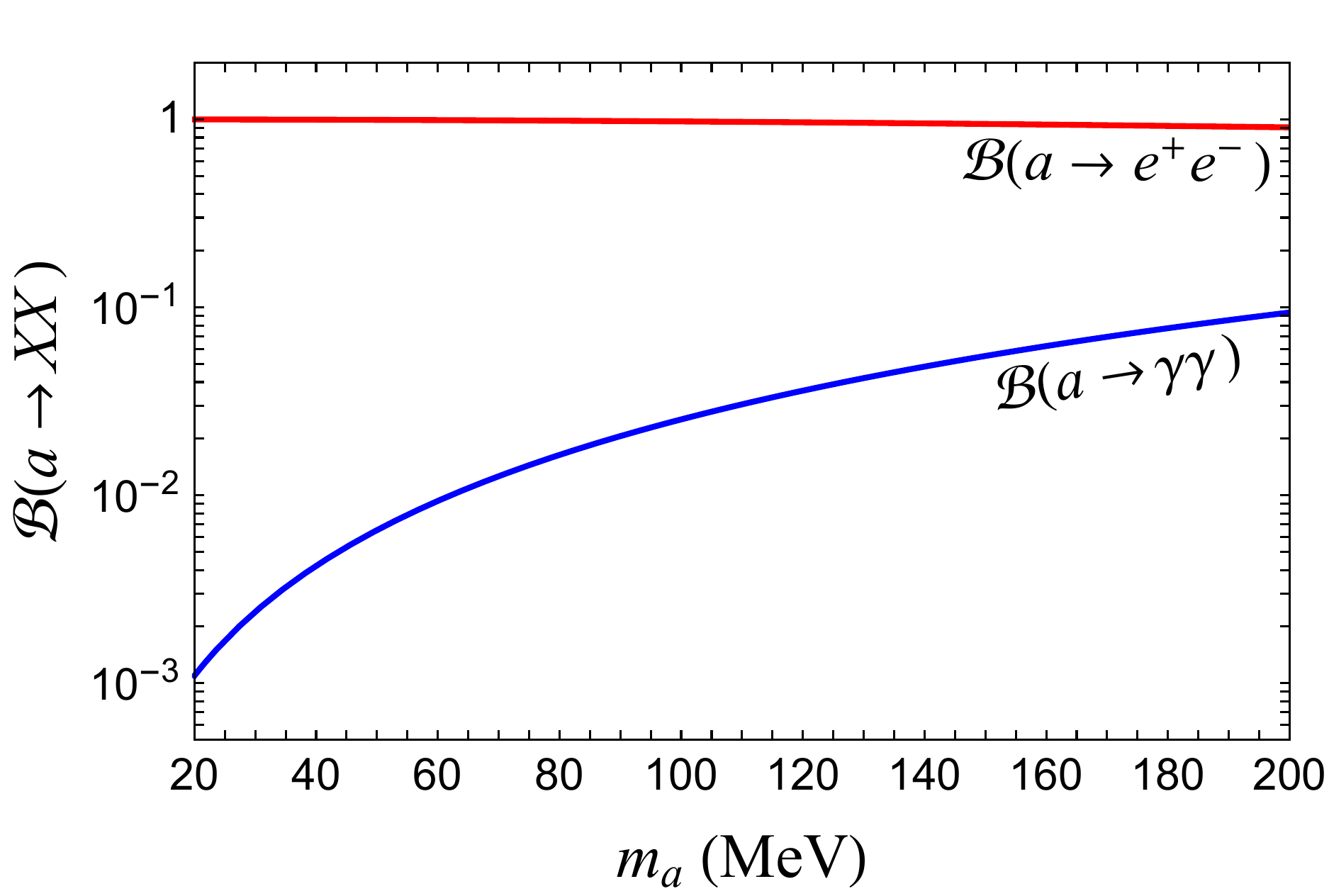}
\caption{The decay branching fractions of the $\ell$ALP, which is independent of $c^{}_e/f^{}_a$.}
\label{fig:BRatoXX}
\end{figure}

Then, the $\ell$ALP can decay into electron-positron and photon pairs with the couplings given in Eqs.\,\eqref{Lint} and \eqref{Lag}.\,\,The decay widths of the $\ell$ALP into $e^+e^-$ and $\gamma\gamma$ are computed, respectively, as
\begin{eqnarray}\label{decayALP}
\Gamma_{a \to e^+e^-} \,=\, \frac{c_e^2 m_e^2 m_a^{}}{8\pi f_a^2}\sqrt{1-\frac{4m_e^2}{m^2_a}} 
~,\quad
\Gamma_{a \to \gamma\gamma} \,=\, \frac{g_{a\gamma\gamma}^2 m_a^3}{64\pi}                  
~.
\end{eqnarray}
We show in Fig.\,\ref{fig:BRatoXX} the decay branching ratios of the $\ell$ALP. As indicated, the $\ell$ALP mainly decays into $e^+e^-$ in the mass range we are interested in.\,\,The decay products $e^+e^-$ inside the detector, we propose, could possibly account for the excess reported by MiniBooNE. Notice that the $m_a^3$ dependence in the $\Gamma_{a \to \gamma\gamma}$ can counteract the $g_{a\gamma\gamma}$ suppression for heavy $\ell$ALP with $m^{}_a \gtrsim 200 \,\text{MeV}$, where the decay channel of $a \to \gamma\gamma$ gives a non-negligible contribution to the total decay width of the $\ell$ALP.

Now, for the sterile neutrino and $\ell$ALP to both decay within the MiniBooNE detector, we have to examine the mean decay distances $d^{}_{N,{}^{}a}$ of both particles in the laboratory frame.\,\,Using the results in Eqs.\,\eqref{decayN} and \eqref{decayALP}, we obtain
\begin{eqnarray}
d^{}_N 
\,=\, 
\gamma^{}_N  {}^{} \beta^{}_N {}^{} \tau^{}_N
\,\simeq\,
1.14 \times 10^3 \,\text{m} \, 
\bigg(\frac{p^{}_N}{3\,\text{GeV}}\bigg)
\bigg(\frac{|U_{\mu{\rm 4}}|}{1\times 10^{-5}}\bigg)^{\hs{-0.15cm}-2}
\bigg(\frac{m^{}_N}{380\,\text{MeV}}\bigg)^{\hs{-0.15cm}-4}
\bigg(\frac{f^{}_a}{100\,\text{GeV}}\bigg)^{\hs{-0.15cm}2}
~,
\end{eqnarray}
\vs{-0.5cm}
\begin{eqnarray}\label{da}
d^{}_a 
\,=\, 
\gamma^{}_a  \beta^{}_a \tau^{}_a
\,\simeq\,
1.42 \,\text{m} \, 
\bigg(\frac{p^{}_a}{3\,\text{GeV}}\bigg)
\bigg(\frac{m^{}_a}{20\,\text{MeV}}\bigg)^{\hs{-0.15cm}-2}
\bigg(\frac{f^{}_a}{100\,\text{GeV}}\bigg)^{\hs{-0.15cm}2}
~,
\end{eqnarray}
where $\gamma^{}_{N} (\gamma^{}_{a})$, $\beta^{}_N (\beta^{}_{a})$, $\tau^{}_N (\tau^{}_{a})$, and $p^{}_N (p^{}_a)$ are the Lorentz boost factor, speed, lifetime, and momentum of the sterile neutrino ($\ell$ALP), respectively. Therefore, with these fiducial values for the masses and couplings, the sterile neutrino and $\ell$ALP can have the proper mean decay lengths which are consistent with the MiniBooNE experimental setup.

\section{Astrophysical and Experimental Constraints}\label{sec:3}

In this section, we will scrutinize the astrophysical and experimental constraints of $c^{}_e/f^{}_a$ and $U_{\mu4}$ in our $\ell$ALP setup.\,\,For these two parameters, the associated astrophysical bounds come from celestial objects such as a red giant, white dwarf, and supernova (SN), depending on the mass scale of the $\ell$ALP. In our work, we will consider the $\ell$ALP with a few tens MeV mass, so the strongest limit is set by the SN1987A. On the other hand, there are several terrestrial laboratories which can place constraints on these parameters as well, including the electron beam-dump experiment E137, electron magnetic dipole moment anomaly, rare kaon decays, and so on.\,\,Finally, for an electron-positron pair to mimic a single electronlike event, we have to demand that the opening angle of an electron-positron pair is small enough.\,\,This would truncate the momentum of the $\ell$ALP at a certain value and then set a lower bound on $c^{}_e/f^{}_a$.\,\,Here, we briefly discuss these constraints in the following.

\subsection{Supernova 1987A}\label{subsec:2}

The observed neutrino burst duration of SN1987A can impose a constraint for tens of MeV axionlike particles (ALPs)\,\cite{Raffelt:1990yz}.\,\,This is because the temperature of the protoneutron star (PNS) can reach of the order of $30\,\text{MeV}$.\,\,With these temperatures, the ALPs can be produced inside the PNS and then carry away a lot of energy from it (which is known as the free-streaming regime).\,\,This process would speed up the cooling rate of the PNS and shrink the period of neutrino burst.\,\,Since the energy loss rate due to the ALP should not exceed the ones via the neutrinos, an approximate analytic bound on the energy loss rate through the ALP in the free-streaming regime is given by~\cite{Raffelt:2006cw}
\begin{eqnarray}\label{Eabound}
\dot{E}^{}_{a} \,\lesssim\, 10^{19} \, \text{erg}\,\text{g}^{-1}\,\text{s}^{-1} ~,
\end{eqnarray}
which is evaluated at the typical core density of $3 \times 10^{14}\,\text{g}\,\text{cm}^{-3}$ and temperature of $30\,\text{MeV}$. 
It is worth mentioning that several numerical simulations demonstrated that the neutrino burst duration would be roughly reduced  by half when the limit of Eq.\,\eqref{Eabound} is saturated~\cite{Raffelt:2006cw}.

Given the couplings in Eqs.\,\eqref{Lint} and \eqref{Lag}, the primary production channels of the ALP inside the PNS are the electron-nucleus bremsstrahlung $(e^- + {\cal N} \to e^- + {\cal N} + a)$ and the Primakoff process $(\gamma + {\cal N} \to {\cal N} + a)$, respectively.\,\,For an ALP with mass of a few tens MeV, it has been estimated in Ref.\,\cite{Calibbi:2020jvd} that the energy loss rate due to the former one is
\begin{eqnarray}\label{EeNeNa}
\dot{E}^{}_{e {\cal N} \to e {\cal N} a} \,\simeq\,
2.84 \times 10^{33} \bigg(\frac{c^{}_e/f^{}_a}{\text{GeV}^{-1}}\bigg)^{\hs{-0.15cm}2} \,\, \text{erg}\,\text{g}^{-1}\,\text{s}^{-1} ~,
\end{eqnarray}
where the Boltzmann suppression factor has been taken into account.\,\,On the other hand, with the help of Eq.\,\eqref{Eabound} and the numerical results analyzed in Refs.\,\cite{Lee:2018lcj,Lucente:2020whw}, the energy loss rate due to the Primakoff process is estimated as
\begin{eqnarray}
\dot{E}^{}_{\gamma {\cal N} \to {\cal N}a} \,\simeq\,
2.78 \times 10^{35} \bigg(\frac{g^{}_{a\gamma\gamma}}{\text{GeV}^{-1}}\bigg)^{\hs{-0.15cm}2} \,\, \text{erg}\,\text{g}^{-1}\,\text{s}^{-1} ~.
\end{eqnarray}
Then, as emphasized below Eq.\,\eqref{Ftau}, it follows that $\dot{E}^{}_{e {\cal N} \to e {\cal N} a}{}^{}/\dot{E}^{}_{\gamma {\cal N} \to {\cal N}a}  \simeq {\cal O}(10^{3})$.\,\,Hence, the energy loss of the PNS is mainly through the electron-nucleus bremsstrahlung in this model. Imposing Eq.\,\eqref{Eabound} to Eq.\,\eqref{EeNeNa}, $\dot{E}^{}_{e {\cal N} \to e {\cal N} a}  \lesssim 10^{19} \, \text{erg}\,\text{g}^{-1}\,\text{s}^{-1}$, we yield 
\begin{eqnarray}
c^{}_e/f^{}_a \, \lesssim \, 6 \times 10^{-8} \,\text{GeV}^{-1}  \quad (\text{free-streaming regime}) ~.
\end{eqnarray}
Note that the above upper bound is no longer valid for sufficiently large $c^{}_e/f^{}_a$.\,\,The reason is that when the coupling strength of the ALP becomes too strong it would be captured within the PNS and cannot escape from it.\,\,This is the so-called trapping regime.\,\,To find out the lower bound of the trapping regime, one can require that the mean free path of the ALP is smaller than the effective radius of the PNS. Following Refs.\,\cite{Raffelt:1990yz,Lucente:2020whw}, the resultant lower bound of $c^{}_e/f^{}_a$ is derived as
\begin{eqnarray}
c^{}_e/f^{}_a \, \gtrsim \, 3 \times 10^{-6} \,\text{GeV}^{-1}  \quad (\text{trapping regime}) ~.
\end{eqnarray}
We present the SN1987A excluded range of $c^{}_e/f^{}_a$ in the yellow shaded region of Fig.\,\ref{fig:cefa_vs_ma}. 

The SN1987A can also give constraints for models with the sterile neutrino.\,\,For example, it has been considered in Ref.\,\cite{Fischer:2019fbw} to restrict $U_{\mu4}$ as the sterile neutrino can be produced in the PNS by the Primakoff upscattering via the photon exchange with nucleons $(\nu + {\cal N} \to {\cal N} + N^{}_\textsf{D})$~\cite{Magill:2018jla}. However, since we assume the ALP in our model is leptophilic.\,\,Thus, it cannot be produced via the $\ell$ALP exchange with nucleons.\,\,Namely, we can evade the constraint of $U_{\mu4}$ from the SN1987A.

\subsection{Electron beam-dump experiment E137}\label{subsec:2}

There exist some experiments searching for long-lived light particles by impinging high-intensity proton or electron beams on the heavy materials, called beam-dump experiments. In the $\ell$ALP model, only the electron beam-dump experiment such as E137~\cite{Bjorken:1988as} is relevant to our study since the $\ell$ALP only interacts with electron and photon.\,\,In the E137 experiment, a 20\,GeV electron beam collides with plates of aluminum immersed in cooling water.\,\,With a large number of electrons (approximately $2\times 10^{20}$) cumulatively hitting on the target, many $\ell$ALPs can be produced from it.\,\,Once the $\ell$ALPs are generated, they would first penetrate a shielding about 179 m, and then reach an open-air decay region 204 m long.\,\,At the end of the decay region, there is a detector which can receive visible signals from the $\ell$ALP decays. 

The main production mechanisms of the $\ell$ALP in the E137 experiment are the Primakoff effect and bremsstrahlung from
electrons~\cite{Bjorken:1988as}, and as pointed out in Fig.\,\ref{fig:BRatoXX}, the $\ell$ALP decays preferentially into an electron-positron pair.\,\,Notice that, although $g^{}_{a\gamma\gamma}$ is much smaller than $c^{}_e/f^{}_a$ in the $\ell$ALP model; however, these two production mechanisms are comparable.\,\,To see this, one can compare the scaling of the cross section of these processes.\,\,In the former case, we have $\sigma_{a\gamma} \propto \alpha {}^{} g^2_{a\gamma\gamma}$, while for the latter one, we have $\sigma_{ae} \propto \alpha^2 {}^{} (m_e/m_a)^2(c^{}_e/f^{}_a)^2$\,\cite{Darme:2020sjf}.\,\,Then, by taking the ratio of them, we find that $\sigma_{ae}/\sigma_{a\gamma} \sim {\cal O}(1)$.

The constraint of $c^{}_e/f^{}_a$ by the E137 experiment is shown in the orange shaded region of Fig.\,\ref{fig:cefa_vs_ma},
where the upper bound and lower bound correspond to the short-lived and long-lived $\ell$ALP, respectively.\,\,In this region, no event has been seen by E137~\cite{Essig:2010gu}.

\subsection{Anomalous electron magnetic dipole moment}\label{subsec:3}

With the $\ell$ALP-electron interaction, there is a one-loop Feynman diagram involving the $\ell$ALP, which contributes to the anomalous electron magnetic dipole moment, $(g - 2)_e$.\,\,The latest measurement deviating from the SM value is given by~\cite{Aoyama:2017uqe,Parker:2018vye}
\begin{eqnarray}\label{Delta_a}
\Delta a^{}_e \,=\, a^{\rm exp}_e - a^{\rm SM}_e  \,=\,  - (8.8 \pm 3.6) \times 10^{-13} ~,
\end{eqnarray}
corresponding to about $2.4^{}\sigma$ tension with the SM prediction, where $a^{}_e \equiv (g - 2)_e/2$. On the other hand, the leading-order contribution to $\Delta a^{}_e$ by the $\ell$ALP can be found in Ref.\,\cite{Abu-Ajamieh:2018ciu} as
\begin{eqnarray}\label{Delta_a_ALP} 
\Delta a^{\ell{\rm ALP}}_e 
\,=\, 
-{}^{}\frac{c_e^2m_e^2}{8\pi^2f_a^2}
\mathop{\mathlarger{\int}_0^1} \hs{-0.08cm} dx \,
\frac{x^3}{x^2 +(1-x) {}^{} r} ~,
\end{eqnarray}
where $r = m_a^2/m_e^2$.\,\,Notice that the $\ell$ALP-photon interaction also gives a contribution to $\Delta a^{}_e$ through the Barr-Zee diagram.\,\,However, this contribution is subleading since $g^{}_{a\gamma\gamma} \ll c^{}_e/f^{}_a$.
Imposing Eq.\,\eqref{Delta_a_ALP} to Eq.\,\eqref{Delta_a}, we find that only the large values of $c^{}_e/f^{}_a$ are subject to the $(g - 2)_e$ anomaly\footnote{A more recent measurement of the $(g-2)_e$ anomaly can be found in Ref.\,\cite{Morel:2020dww}, where $\Delta a^{}_e \,=\, (4.8 \pm 3.0) \times 10^{-13}\,(+1.6\sigma)$. In this case, Eq.\,\eqref{Delta_a_ALP} can only contribute to the negative error bar, and it does not affect our result too much.}; see the green shaded region in Fig.\,\ref{fig:cefa_vs_ma}.

\subsection{Collimated $e^+e^-$ pair as a single electronlike event}\label{subsec:4}
\begin{figure}[t!]
\vs{-0.3cm}
\hs{-1.0cm}
\centering
\includegraphics[scale=0.45]{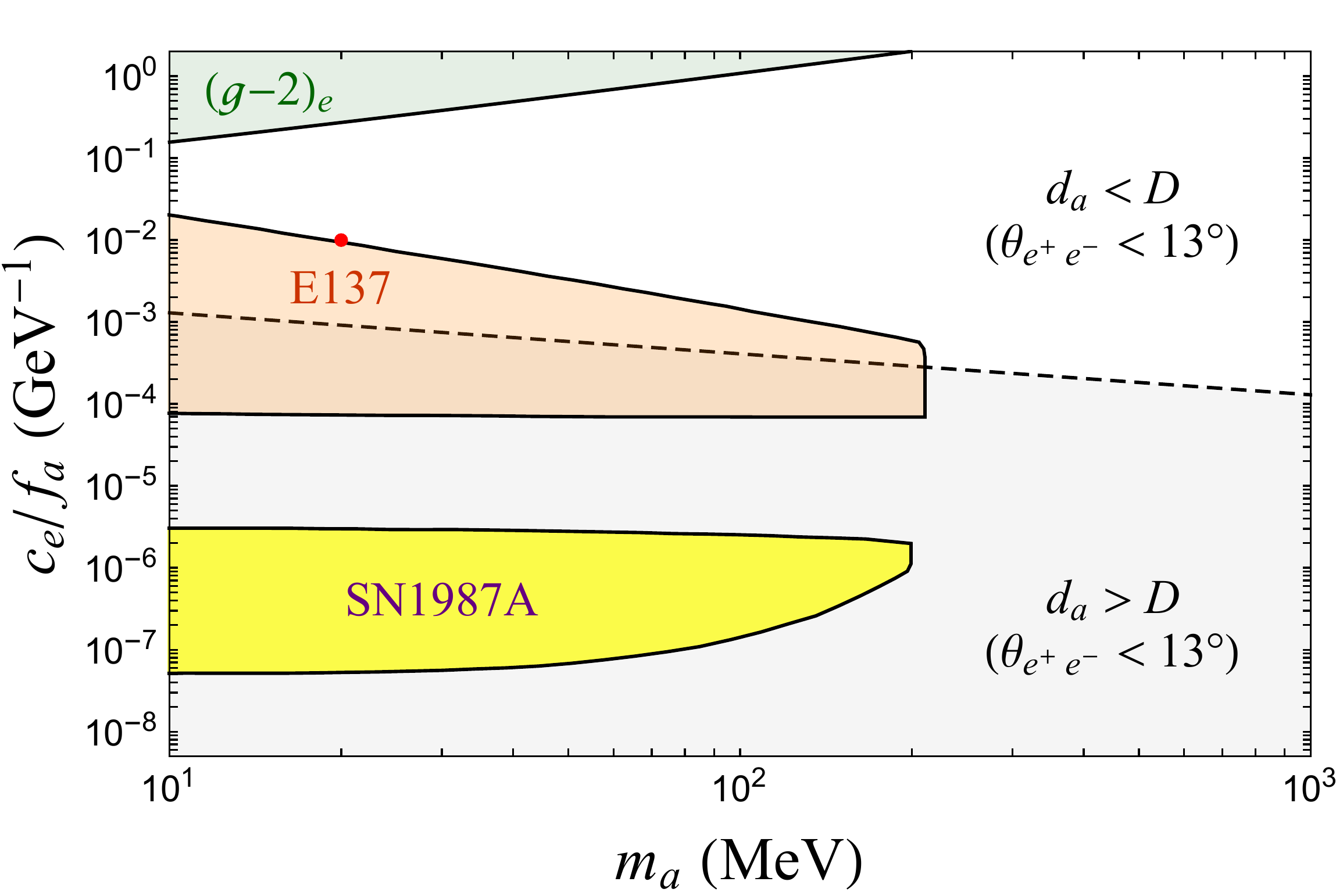}
\caption{The various astrophysical and experimental bounds of $c^{}_e/f^{}_a$ in the $\ell$ALP model, where the yellow, orange, and green shaded regions are excluded by the SN1987A, electron beam-dump experiment E137, and $(g-2)^{}_e$ anomaly, respectively. In the gray shaded area, the number of events is suppressed as the $\ell$ALP mostly decays outside the MiniBooNE detector.\,\,Our benchmark point is indicated by the red dot right above the upper boundary of the E137 constraint.}
\label{fig:cefa_vs_ma}
\end{figure}

For an electron-positron pair produced from the $\ell$ALP decay been identified with an electronlike signal, we have to require that the opening angle of an electron-positron pair is $\theta_{e^+e^-} \hs{-0.05cm} < 13$ deg~\cite{Aguilar-Arevalo:2018gpe}.\,\,This upper bound of the opening angle can be translated into the lower bound of the momentum of the $\ell$ALP as~\cite{Amsler:2007cyq}
\begin{eqnarray}\label{pamin}
p^{}_a \,>\,p^{}_{a,\text{min}} \,\equiv\, \sqrt{m^2_a-4m_e^2} \, \cot\hs{-0.05cm}\big(\theta_{e^+e^-}/2\big) 
\,\,\simeq\,\, 8.78\,m^{}_a
~.\quad
\end{eqnarray}
Now, for an electron-positron pair to be able to be detected in the MiniBooNE experiment, the mean decay length of the $\ell$ALP should be smaller than the diameter of the MiniBooNE detector.\,\,Using Eq.\,\eqref{da} with Eq.\,\eqref{pamin} and requiring $d^{}_a < D =10\,\text{m}$, we arrive at
\begin{eqnarray}\label{damin}
c^{}_e/f^{}_a \,\gtrsim\,
9.13 \times10^{-4} \,\text{GeV}^{-1} \bigg(\frac{m^{}_a}{20\,\text{MeV}}\bigg)^{\hs{-0.15cm}-1/2} ~.
\end{eqnarray}
Note that Eq.\,\eqref{damin} is a conservative bound, below which the $\ell$ALP still has a probability to decay within the detector.\,\,This factor will be considered in the computation of the excess events in the next section.\,We show this bound as the black dashed line in Fig.\,\ref{fig:cefa_vs_ma}.

\hs{1cm}

Based on the above constraints, we choose an optimistic benchmark point, $m^{}_a = 20 \,\text{MeV}$ and $c^{}_e/f^{}_a = 10^{-2} \,\text{GeV}^{-1}$\,(the red dot in Fig.\,\ref{fig:cefa_vs_ma}) in our numerical calculation.\,\,Also, we have checked that this benchmark point is far below the sensitivities of the current colliders~\cite{Bauer:2017ris} and far above the limits from cosmology~\cite{Cadamuro:2011fd}.

\begin{figure}[htbp]
\hs{-1.5cm}
\centering
\includegraphics[scale=0.5]{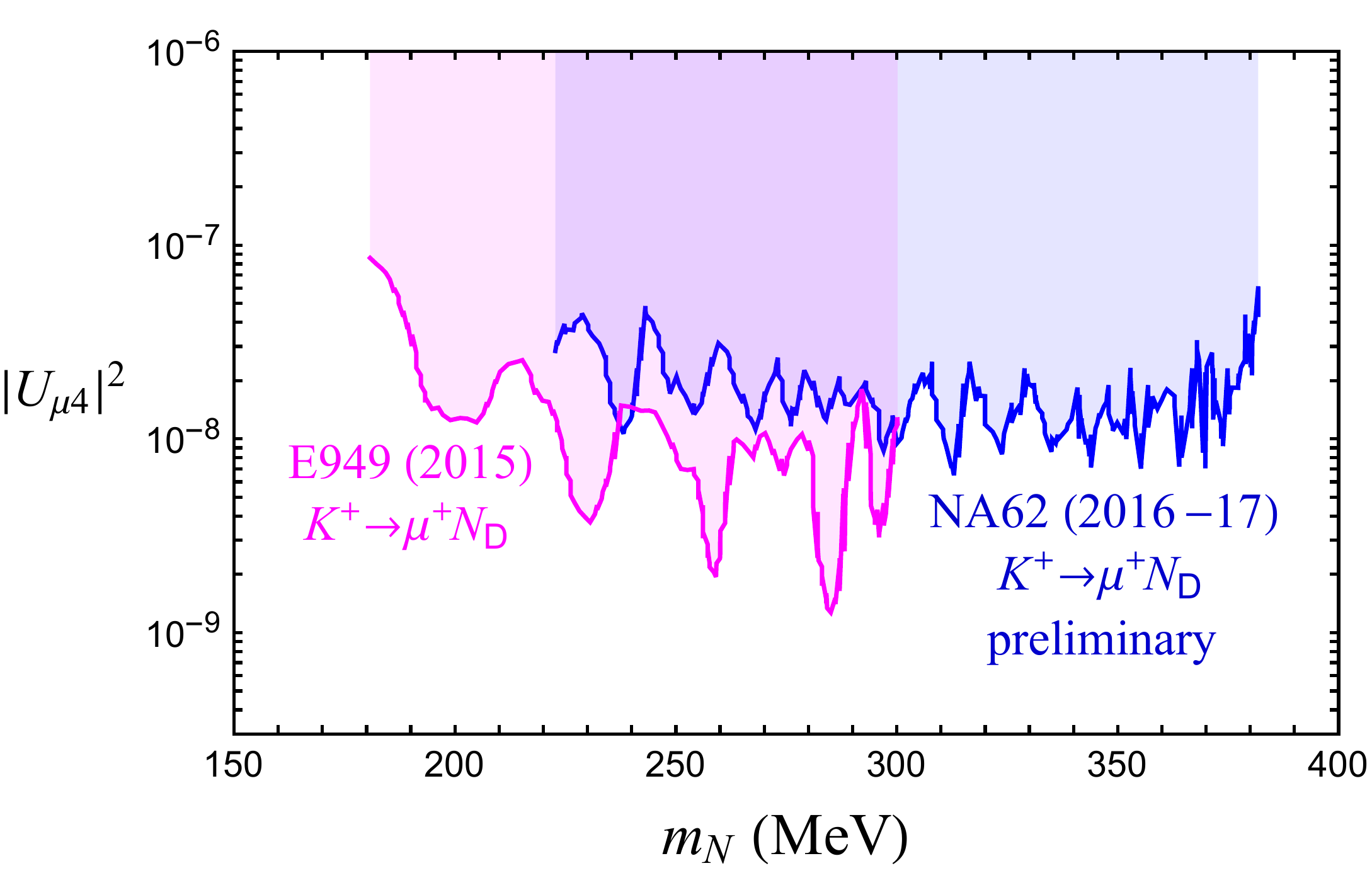}
\caption{The upper limits of $|U_{\mu4}|^2$ from the kaon decay experiments at 90\% C.L., where the magenta and blue shaded regions are excluded by E949 and NA62, respectively.}
\label{fig:Umu4_mN_limit}
\end{figure}

\subsection{Rare kaon decays}\label{subsec:5}
In our setup, the sterile neutrino is produced from the charged kaon decay, say, $K^+ \to \mu^+ N^{}_{\textsf D}$, through the mixing between the sterile neutrino and the muon neutrino.\,\,With the neutrino mixing parameter $U_{\mu 4}$, the corresponding branching fraction is given by\,\cite{CortinaGil:2017mqf}
\begin{eqnarray}\label{BKtoND}
{\cal B}(K^+ \to \mu^+ N^{}_{\textsf D}) 
\,=\,
{\cal B}(K^+ \to \mu^+ \nu_\mu) {}^{} \rho^{}_\mu (m^{}_N) {}^{}  |U_{\mu 4}|^2  ~,
\end{eqnarray}
where ${\cal B}(K^+ \to \mu^+ \nu_\mu) = 0.6356$~\cite{Zyla:2020zbs}, and the kinematical function
\begin{eqnarray}
\rho^{}_\mu(m^{}_N)
\,=\,
\frac{x^{}_N + x^{}_\mu-(x_N^{} - x^{}_\mu)^2}{x^{}_\mu (1-x^{}_\mu)\raisebox{0.5pt}{$^2$}} 
\big[\lambda(1,x^{}_N,x^{}_\mu)\big]^{\hs{-0.05cm}1/2}
\end{eqnarray}
with $x^{}_N \equiv m^2_N/m^2_K$, $x^{}_\mu \equiv m^2_\mu/m^2_K$, and $\lambda(u,v,w) = u^2 + v^2 + w^2 - 2(u^{}v + v^{}w + w^{}u)$. The E949~\cite{E949} and NA62~\cite{NA62} are past and current kaon decay experiments searching for the sterile neutrino through this decay process.\,\,By measuring the muon momentum spectrum or missing energy spectrum of the kaon decays, they can provide upper limits for $|U_{\mu 4}|^2$, which are displayed as color lines in Fig.\,\ref{fig:Umu4_mN_limit}.\,\,In this figure, we take the data of the E949 experiment from Ref.\,\cite{Artamonov:2014urb}, and for the NA62 ones, we adopt the preliminary updated result (approximately $\hs{-0.05cm}1/3$ of the data set) announced in Ref.\,\cite{Goudzovski:NA62}.\,\,As indicated, the E949 experiment places the most stringent upper limit on $|U_{\mu 4}|^2$ down to approximately $10^{-9}$ for the sterile neutrino with masses between 175 and 300\,MeV. On the other hand, the NA62 experiment extends the search range of the sterile neutrino mass from 300 to 383\,MeV, where $|U_{\mu 4}|^2 \lesssim 10^{-8}$.\,\,In the next section, we will take $m^{}_N = 380\,\text{MeV}$ and $|U_{\mu 4}| =1.6\times 10^{-5}$ as the benchmark point for our computations.

\section{MiniBooNE excess events and our fitting results}\label{sec:4}

In this section, we will outline how we compute the excess event numbers in the $\ell$ALP model.\,\,Essentially, we follow the approaches given in Refs.\,\cite{Fischer:2019fbw,Hernandez-Cabezudo:2020weh} with some modifications. They consider a sterile neutrino decaying inside the detector into an active neutrino and a photon.\,\,The work reconstructs first the kaon flux from the given flux of the muon neutrino. From the kaon flux, they then derive the sterile neutrino flux.\,\,We follow similar procedures. In comparison, however, we replace the massless photon with an unstable massive $\ell$ALP, and it is expected that the kinematics in our calculation is a little bit different from theirs. Indeed, the condition of the opening angle of an electron-positron pair by the $\ell$ALP decay requires a minimum of the $\ell$ALP momentum.\,\,In the computation, this may eliminate some of the events from contributing to excesses.\,\,In the following, we will first write down all the relevant formulas for estimating the total number of events and then present our numerical results before the end of this section.

It is worth it to mention that the production of $K^+$\,($\text{or}\,K^-$) at the target can be parametrized using the Feynman scaling~\cite{Feynman:1969ej}.\,\,With the best-fit parameters provided in Ref.\,\cite{Mariani:2011zd}, one can generate the momentum distribution of $K^+$.\,\,Then, it would be straightforward to get the kinematics of sterile neutrino, $\ell$ALP, and electron-positron pair in the decay chain of $K^+\to \mu^+ N^{}_\textsf{D} \to \mu^+ \nu_\mu {}^{} a$, followed by $a\to e^+ e^-$.\,\,Given the total number of $K^+$ being produced at the target, we can estimate the excess of $\nu_e$-like events mimicked by collimated $e^+e^-$ pairs.\,\,We adopt this method as a cross-check and obtain similar results.

\begin{figure}[t!]
\vs{-0.32cm}
\hs{-0.8cm}
\centering
\includegraphics[scale=0.35]{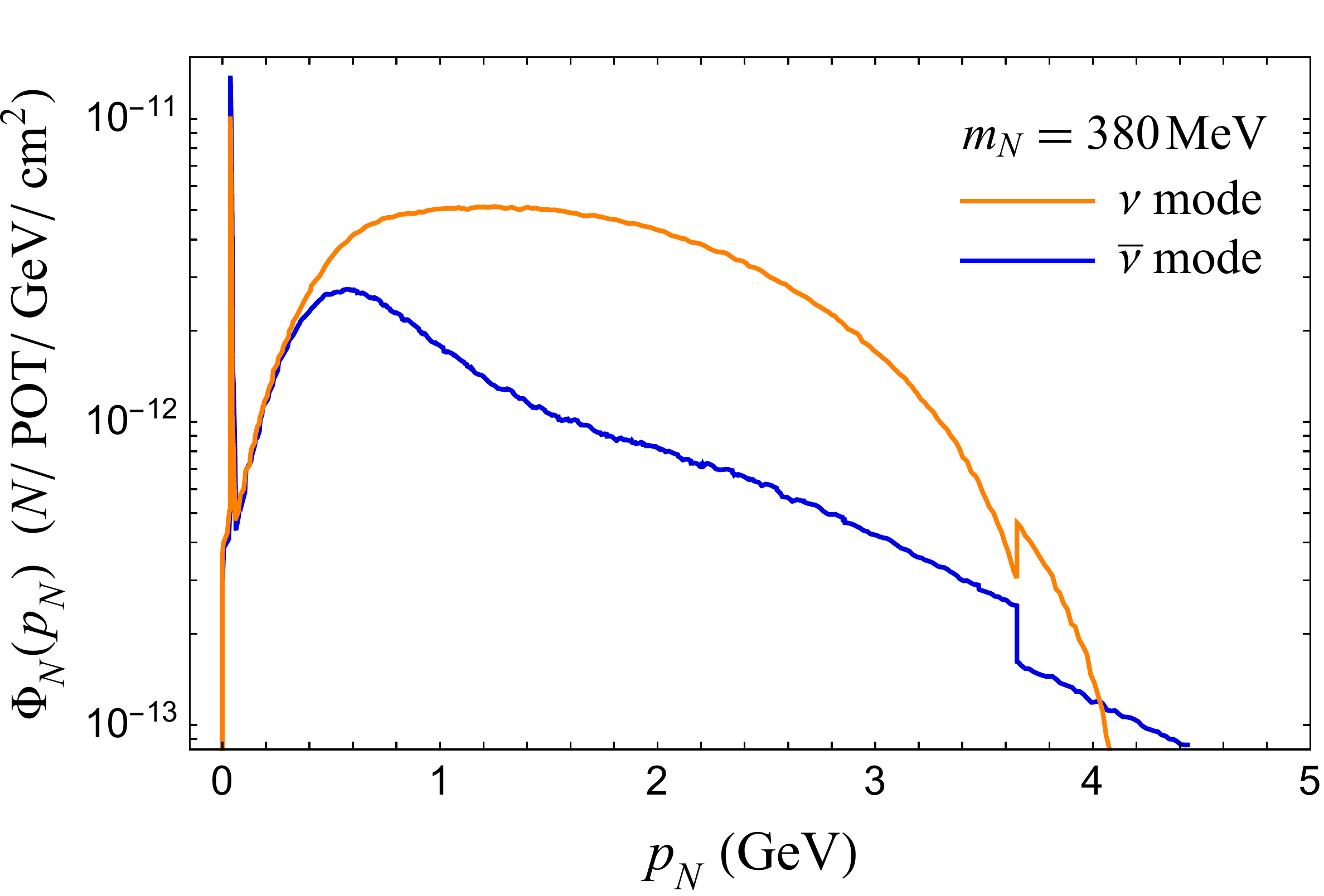}
\caption{The fluxes of the sterile neutrino as a function of $p^{}_N$ for the neutrino (orange line) and antineutrino (blue line) modes with $m^{}_N = 380\,\text{MeV}$, where the sharp peaks correspond to the stopped kaons. Note that $N^{}_\textsf{D}$ and $\bar{N}^{}_\textsf{D}$ equally contribute to $\Phi_N$ as the signal at the MiniBooNE detector does not distinguish between them.}
\label{fig:flux}
\end{figure}

\subsection{Angular and energy spectra of $\ell$ALP}\label{subsec:6}

\begin{figure}[t!]
\hs{-0.15cm}
\centering
\includegraphics[scale=0.462]{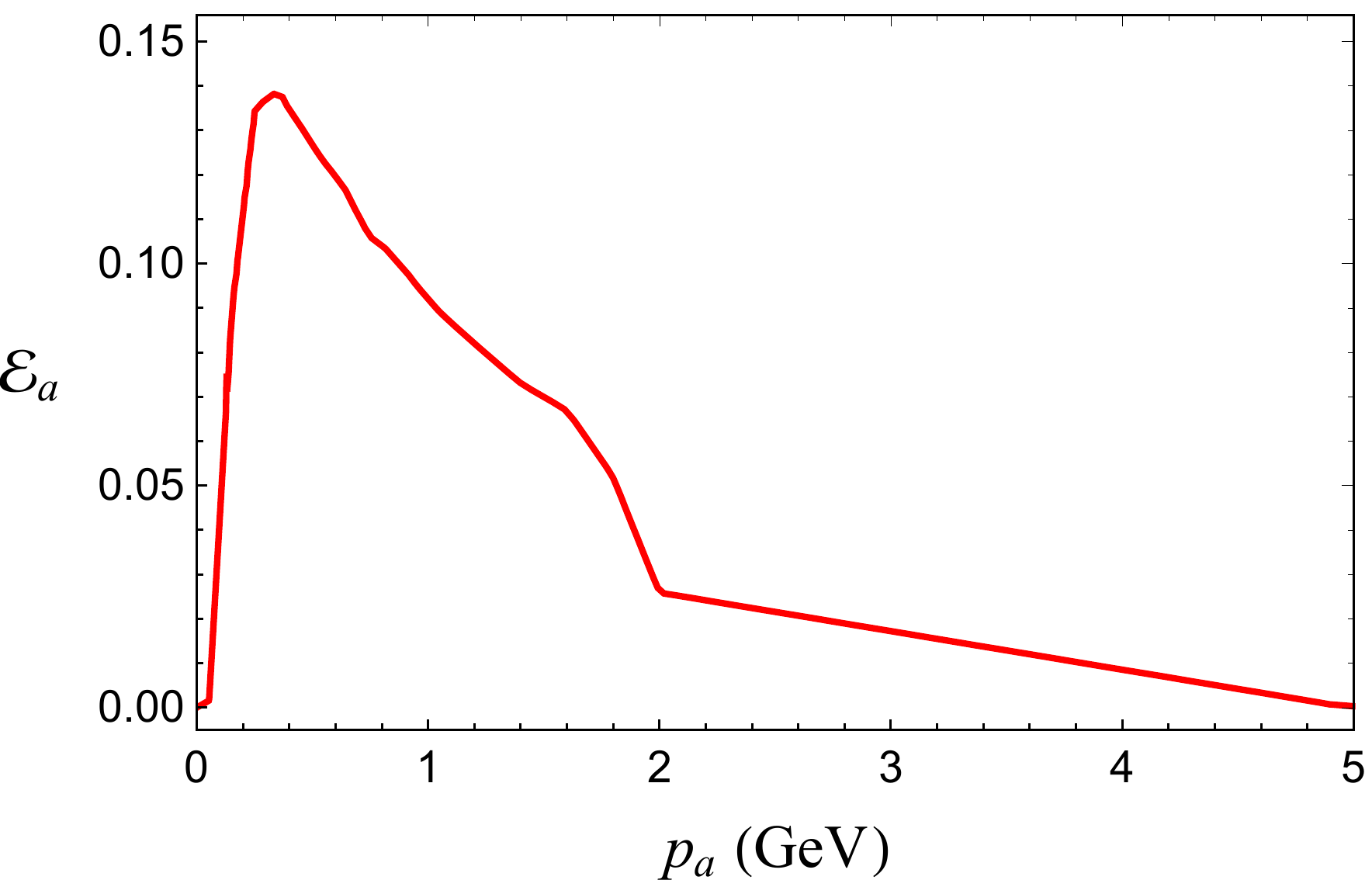}
\hs{0.00cm}
\includegraphics[scale=0.462]{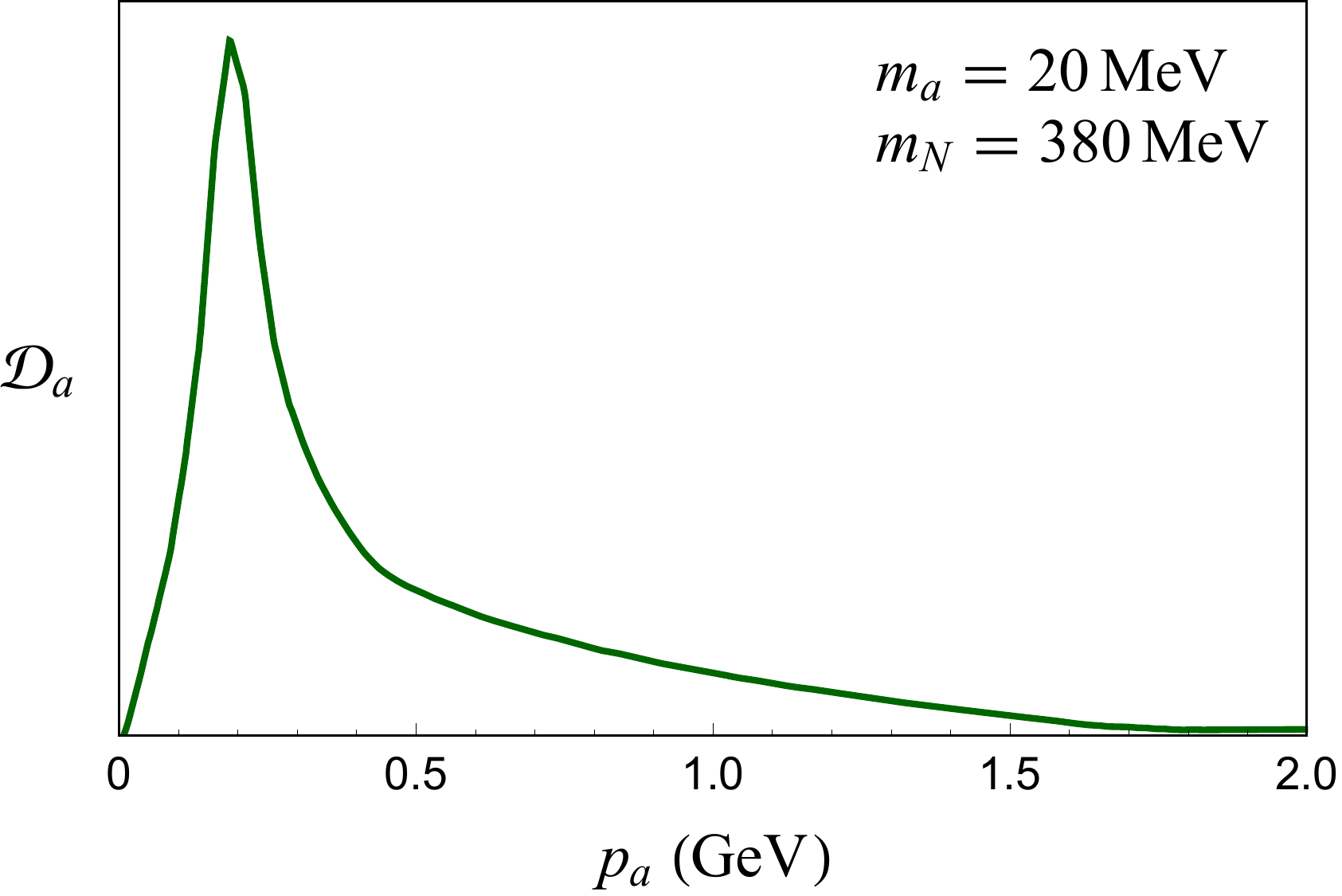}
\caption{Left panel\,\,:\,\,the MiniBooNE detection efficiency of the signal energy, here approximated with the momentum of the $\ell$ALP. Right panel\,\,:\,\,the momentum distribution of the $\ell$ALP with $m^{}_a = 20\,\text{MeV}$ and $m^{}_N = 380\,\text{MeV}$.}
\label{fig:eff_dis}
\end{figure}

In our model, the angle and energy of the signal are interpreted as the scattering angle $\theta_a$ and energy $E^{}_a$ of the $\ell$ALP, which decays into a small opening angle electron-positron pair.\,\,To compute the distributions of the excess events, we integrate the decay spectra of the $\ell$ALP over the sterile neutrino flux, $\Phi^{}_N(p^{}_N)$ (see Fig.\,\ref{fig:flux}), together with the probabilities $P^{}_{N, {\rm dec}}(p^{}_N)$ and $P^{}_{a, {\rm dec}}(p^{}_N)$ that the sterile neutrino and  the $\ell$ALP decay in the MiniBooNE detector, respectively.\,\,Other necessary factors will be explained below.\,\,Since we construct the sterile neutrino flux from the muon neutrino flux of the kaon decay, a normalized factor ${\cal B}(K \to \mu N^{}_{\textsf D})/{\cal B}(K \to \mu {}^{} \nu) = \rho^{}_\mu(m^{}_N) |U_{\mu 4}|^2$ should be included to account for the neutrino mixing $U_{\mu 4}$ and different kinematics of the muon neutrino and heavier sterile neutrino.\,\,The predicted spectrum ${\cal S}$ with respect to the variable $Q\,:\,Q = \cos\theta_a \, {\rm or}  \,  E^{}_a$ can be written as a master formula,
\begin{eqnarray}\label{spectrum}
{\cal S}(Q) 
\Eq
\rho^{}_\mu(m^{}_N) |U_{\mu 4}|^2 \, {\rm POT} {}^{}  A^{}_{\rm MB} 
\mathop{\mathlarger{\int}} \hs{-0.08cm} dp^{}_N {}^{} \Phi^{}_N(p^{}_N) {}^{}{}^{} 
P^{}_{N, {\rm dec}}(p^{}_N) {}^{}{}^{} {\cal W}^{}_{\rm time}(p^{}_N) {}^{}{}^{} 
\nonumber\\
&&\hs{4.7cm}\times\hs{-0.05cm}
\frac{1}{\Gamma^{\rm lab}_N}\frac{d{}^{}\Gamma^{\rm lab}_{N^{}_\textsf{D} \to a\nu}}{d{}^{}Q} {}^{}
P^{}_{a, {\rm dec}}(p^{}_a) {}^{}{}^{} 
{\cal E}^{}_a(p^{}_a) {}^{}{}^{} 
{\cal D}^{}_a(p^{}_a) {}^{}{}^{} 
~,
\end{eqnarray}
where POT denotes the number of protons on target, which is equal to $18.75\,(11.27)\times 10^{20}$ for the neutrino (antineutrino) operation mode~\cite{MiniBooNE:2020pnu}, $A^{}_{\rm MB} = \pi (D/2)^2$ is the effective area of the MiniBooNE detector, ${\cal W}^{}_{\rm time}(p^{}_N) $ is the timing-related weight due to the fact that the sterile neutrino arrives at the detector later than the light ones in a proton beam pulse~\cite{Fischer:2019fbw}
\begin{eqnarray}
{\cal W}^{}_{\rm time}(p^{}_N) 
\,=\, 
{\cal H}\big(\Delta t\big) \frac{\Delta t}{\delta t}  ~,\quad 
\Delta t \,=\, t^{}_0 + \delta t - t^{}_N  ~,\quad 
{\cal H}\big(\Delta t\big)
\,=\,
\begin{cases}
\,\,1  &\text{if} \quad   \Delta t > 0
\\
\,\,0  &\text{if}  \quad  \Delta t < 0
\end{cases}
\end{eqnarray}
with $t^{}_0 = L/c \,\simeq 1.67\,\mu\text{s} \,\, (t^{}_N = t^{}_0/\beta^{}_N)$ being the light (sterile) neutrino arrival time from the source to the detector and $\delta t \,\simeq 1.6\,\mu\text{s}$ being the time interval of the proton beam pulse, and ${\cal E}^{}_a(p^{}_a)$ and ${\cal D}^{}_a(p^{}_a)$ are the MiniBooNE detector efficiency~\cite{Hernandez-Cabezudo:2020weh} and the momentum distribution of the $\ell$ALP as functions of the $\ell$ALP momentum, respectively, which are displayed in Fig.\,\ref{fig:eff_dis}. Note that ${\cal D}^{}_a(p^{}_a)$ has to be normalized when performing the integral in Eq.\,\eqref{spectrum}.

For the probabilities of the sterile neutrino and the $\ell$ALP decaying inside the detectable region of the MiniBooNE experiment, we have
\begin{eqnarray}\label{PNdec}
P^{}_{N,{\rm dec}}(p^{}_N) 
\,=\,
\exp\bigg(\hs{-0.08cm}{-}{}^{}L {}^{}\Gamma^{}_N \frac{m^{}_N}{p^{}_N}{}^{}\bigg) 
\bigg[1-
\exp\bigg(\hs{-0.08cm}{-}{}^{}D{}^{}\Gamma^{}_N \frac{m^{}_N}{p^{}_N}{}^{}\bigg)
\bigg] ~,
\end{eqnarray}
\begin{eqnarray}
P^{}_{a,{\rm dec}}(p^{}_a) 
\,=\, 
1 - \exp\bigg(\hs{-0.08cm}{-}{}^{}D {}^{} \Gamma^{}_a \frac{m^{}_a}{p^{}_a}{}^{}\bigg) ~,\quad
\end{eqnarray}
where $\Gamma^{}_N \simeq \Gamma^{}_{N^{}_\textsf{D} \to a {}^{} \nu}$ and $\Gamma^{}_a \simeq \Gamma^{}_{a\to e^+e^-}$. Now, in the case of the angular spectrum, ${\cal S}(\cos\theta_a)$, the normalized differential decay rate of the sterile neutrino with respect to $\cos\theta_a$ in the laboratory frame is derived as
\begin{eqnarray}
\frac{1}{\Gamma^{\rm lab}_N}\frac{d{}^{}\Gamma^{\rm lab}_{N^{}_\textsf{D} \to a\nu}}{d{}^{}\text{cos}\,\theta_a} \,=\,
\frac{1}{1-m_a^2/m_N^2}\frac{p_a^2}{|^{}p^{}_a E^{}_N - p^{}_N E^{}_a \cos\theta_a|} ~,
\end{eqnarray}
where the $\ell$ALP momentum is given by\footnote{There is an ALP momentum conjugated to $\,p^{}_a$, which is kinematically allowed in the laboratory frame.\,\,However, we have checked that this conjugate momentum is always far below $p^{}_{a,{\rm min}}$; then, it would not contribute to our calculation.}\,\cite{Schlippe:2002}
\begin{eqnarray}\label{pa}
p^{}_a 
\,=\,
\frac{\big(m_N^2+m_a^2\big) p^{}_N \cos\theta_a + E^{}_N \sqrt{\big(m^2_N-m^2_a\big)
\raisebox{0.5pt}{$^{\hs{-0.03cm}2}$} - 4 {}^{} m_a^2 {}^{}{}^{} p^2_N {}^{}{}^{} \text{sin}^2\theta_a}}
{2\big(m_N^2 + p^2_N {}^{}{}^{} \text{sin}^2\theta_a\big)} ~.
\end{eqnarray}
For the energy spectrum, ${\cal S}(E^{}_a)$, one can use the chain rule and \eqref{pa} to derive $d{}^{}\Gamma^{\rm lab}_{N^{}_\textsf{D} \to a\nu}/dE^{}_a$. Note that what is reported by the MiniBooNe experiment is the spectrum of the visible energy, $E^{}_\text{vis}$ (or the reconstructed neutrino energy, $E^{\rm rec}_\nu$).\,\,Since the $\ell$ALP in our model decays visibly, we can approximately take $E^{}_a \approx E^{}_\text{vis}$. 

With the above tools, we can then compute the excess event numbers ${\cal N}^{}_{\cos\theta_a,{}^{}i}$ and ${\cal N}^{}_{E^{}_a,{}^{}i} $ of $i$th bin of the $\ell$ALP angular and energy spectra, respectively, as
\begin{eqnarray}\label{events}
{\cal N}^{}_{\cos\theta_a,{}^{}i} 
\,=\,
\mathop{\mathlarger{\int}_{\cos\theta_{a,i}}^{\,\cos\theta_{a,i+1}} } \hs{-0.08cm} d{}^{}\text{cos}\,\theta_a \, {\cal S}(\cos\theta_a) 
~,\quad
{\cal N}^{}_{E^{}_a,{}^{}i} 
\,=\,
\mathop{\mathlarger{\int}_{E^{}_{a,i}}^{E^{}_{a,i+1}} } \hs{-0.08cm} dE^{}_a \, {\cal S}(E^{}_a)  ~,
\end{eqnarray}
and the total events can be obtained easily by summing up the event numbers in each bin
\begin{eqnarray}\label{totalevents}
{\cal N}^{}_{\cos\theta_a,{}^{}\text{total}} 
\,=\, \sum_i \, {\cal N}^{}_{\cos\theta_a,{}^{}i} 
~,\quad
{\cal N}^{}_{E^{}_a,{}^{}\text{total}} 
\,=\, \sum_i \, {\cal N}^{}_{E^{}_a,{}^{}i}  ~.
\end{eqnarray}
Note that, when evaluating the integrals in \eqref{events}, the cut of the $\ell$ALP momentum, $p^{}_a > p^{}_{a,\text{min}}$, must be considered; see Eq.\,\eqref{pamin}.

\begin{figure}[t!]
\vs{-0.2cm}
\centering
\includegraphics[scale=0.483]{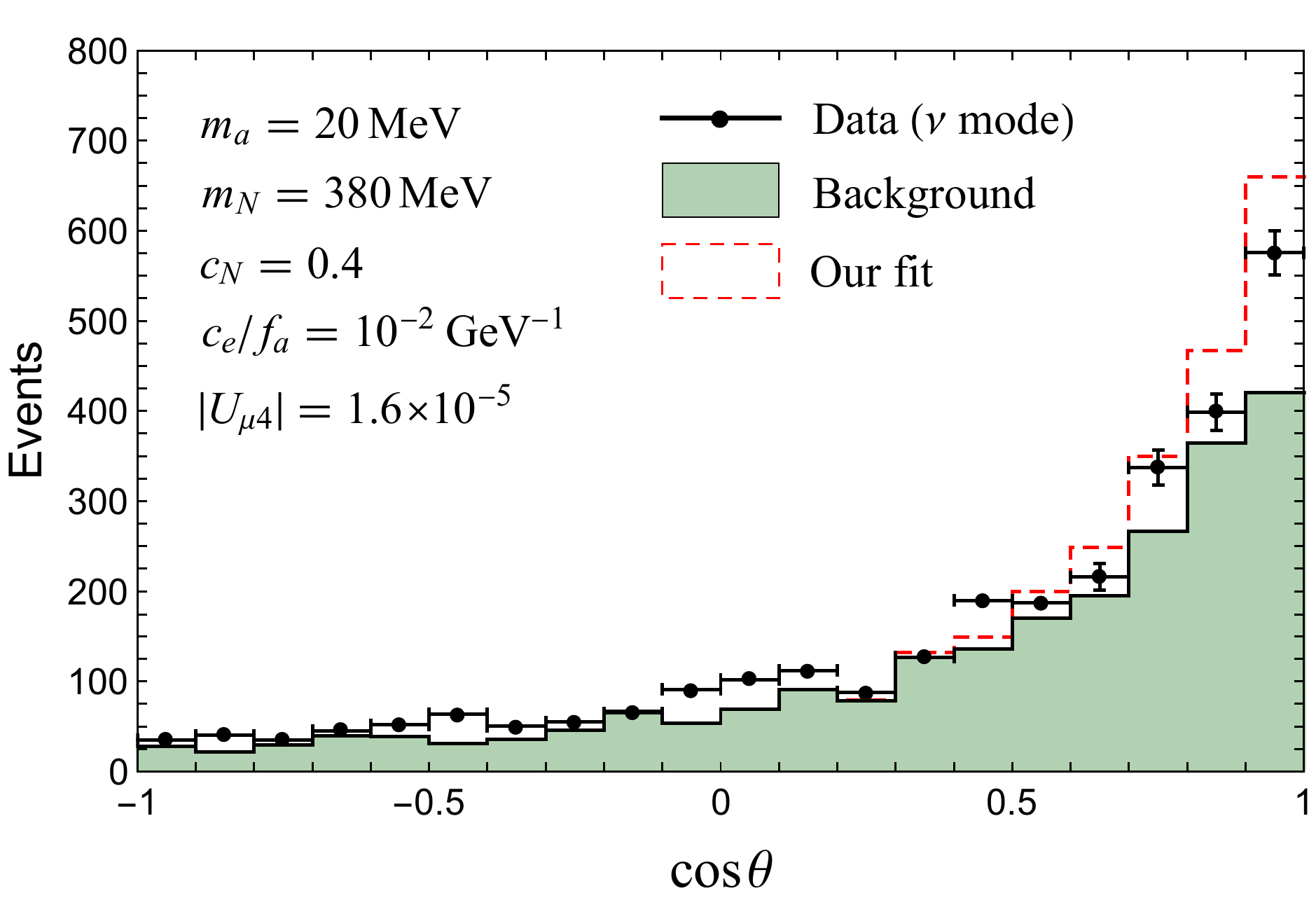}
\\[0.3cm]
\includegraphics[scale=0.48]{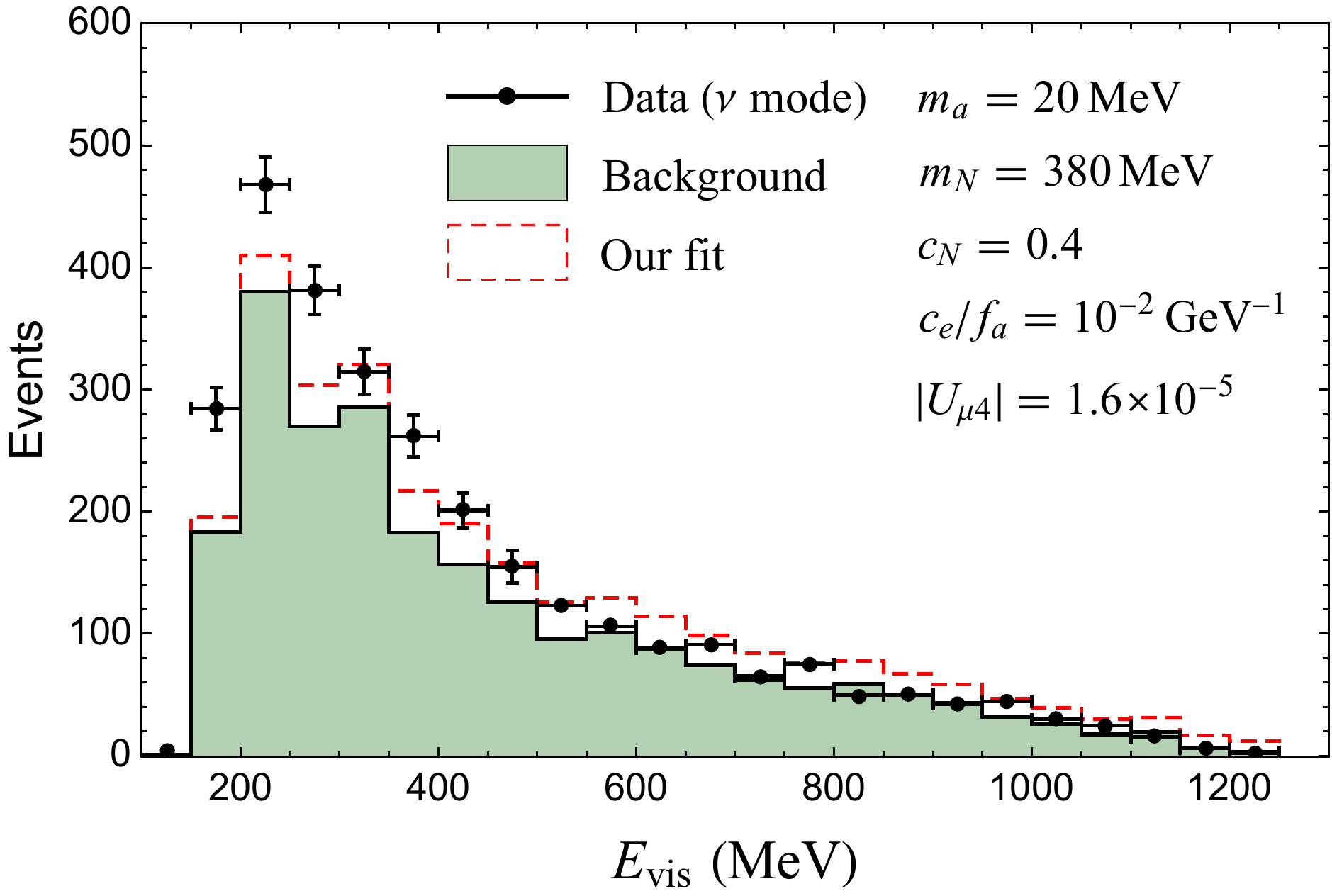}
\caption{Our numerical results for the angular spectrum (top panel) and visible energy spectrum (bottom panel) of the MiniBooNE experiment in the neutrino mode, where the black dots are the excess electronlike events with errors and the green shaded region is the estimated backgrounds. With the benchmark point, the corresponding fittings are shown as red dashed lines in the figures.}
\label{fig:fitting}
\end{figure}

\begin{figure}[t!]
\vs{-0.2cm}
\hs{-1.5cm}
\centering
\includegraphics[scale=0.5]{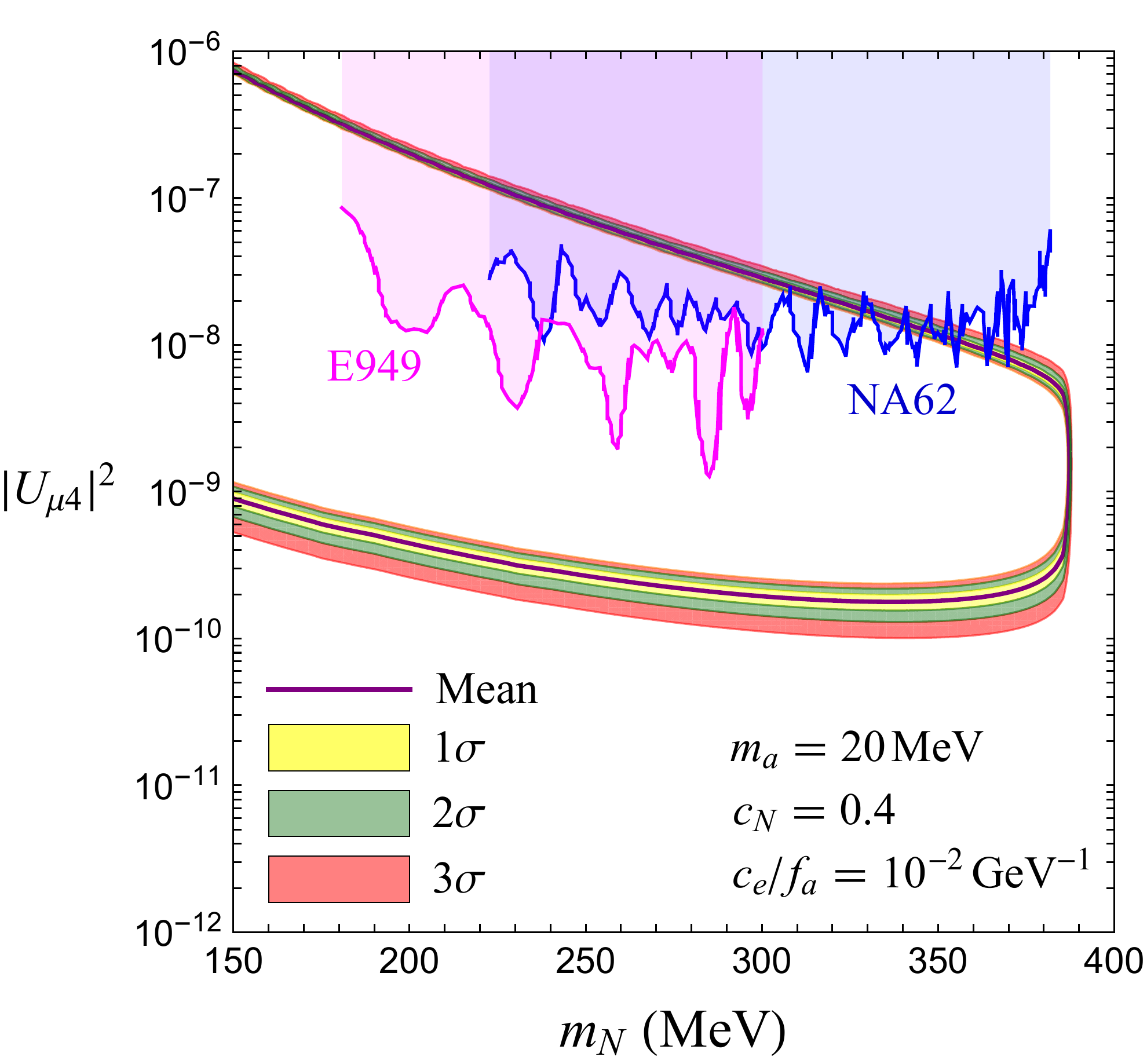}
\\
\centering
\hs{-1.4cm}
\includegraphics[scale=0.505]{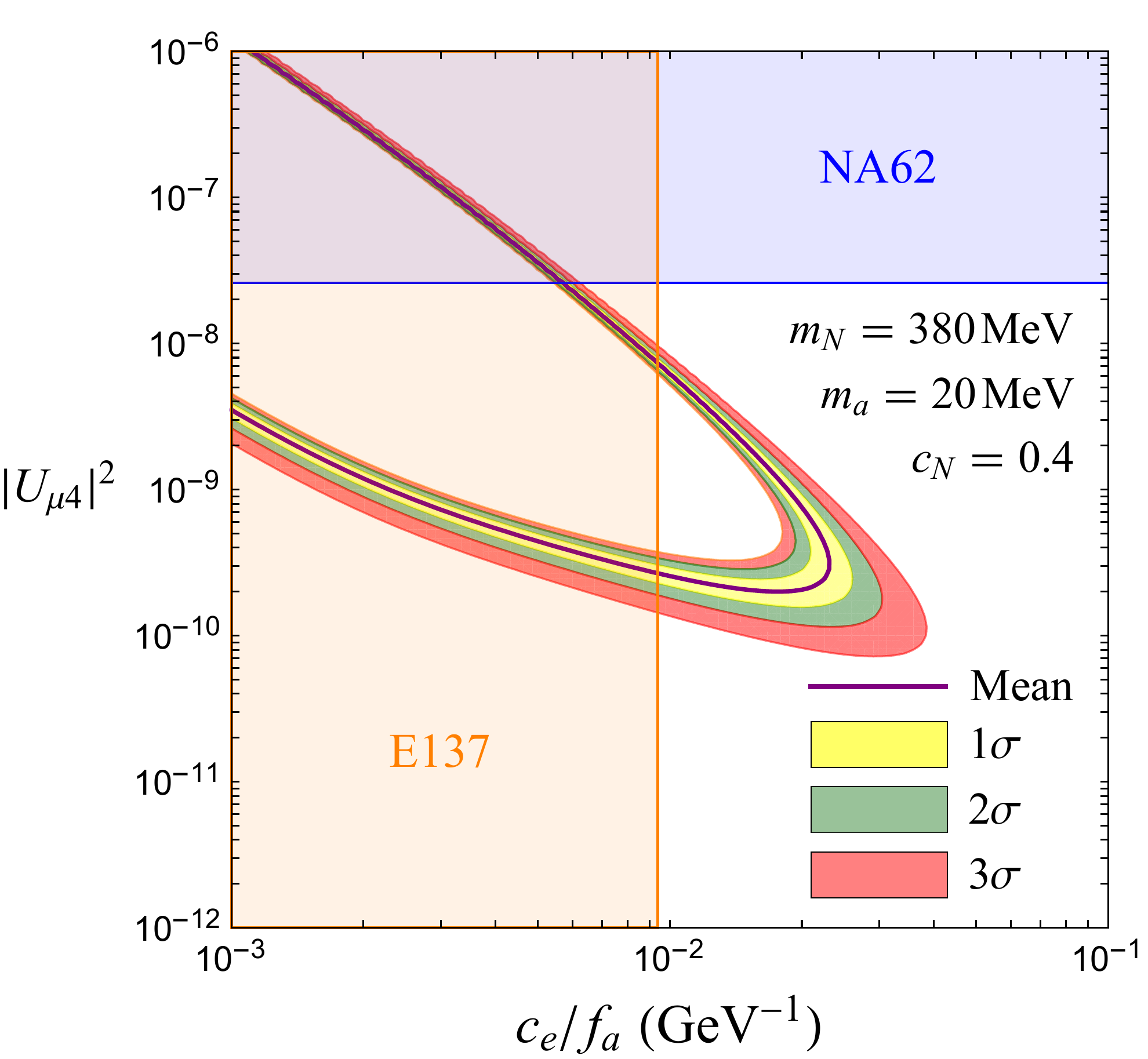}
\caption{The parameter space of the $|U_{\mu 4}|^2$ vs $m^{}_N$ plane (top panel) and $|U_{\mu 4}|^2$ vs $c^{}_e/f^{}_a$ plane (bottom panel) satisfying the MiniBooNE data at $1^{}\sigma$ to $3^{}\sigma$ C.L. in the $\ell$ALP model, where the purple contour presents the mean value of the data.\,\,The shaded regions above the magenta and blue lines are excluded by the kaon decay experiments, E949 and NA62, respectively, and the shaded area left to the orange line is disfavored by the E137 experiment.}
\label{fig:Umu4_mN_fa}
\end{figure}

\subsection{Our fitting results}\label{subsec:7}
Applying the formulas from Eq.\,\eqref{spectrum} to Eq.\,\eqref{events}, we present in Fig.\,\ref{fig:fitting} our fitting results of the angular and visible energy spectra in the neutrino mode, including all of the excess data and expected backgrounds reported by the most recent update analysis of MiniBooNE~\cite{MiniBooNE:2020pnu}.\,\,In both figures, we assume $m^{}_a = 20\,\text{MeV},\,m^{}_N = 380\,\text{MeV},\,c^{}_N =0.4,\,c^{}_e/f^{}_a = 10^{-2}\,\text{GeV}^{-1}$, and $|U_{\mu4}| =1.6 \times 10^{-5}$ as the benchmark point.\,\,One can see that our predictions of the spectra in the $\ell$ALP model are consistent with the tendencies of the experimental data points.\,\,Also, the predicted total excess events are within the $1{}^{}\sigma$ range of the observed ones.\,\,The fitting result of the reconstructed neutrino energy spectrum is similar to that of the visible energy spectrum and is not shown here.

We then use Eq.\,\eqref{totalevents} to calculate the total excess events and depict the allowed regions of parameter space that can explain MiniBooNE data.\,\,At the top panel of Fig.\,\ref{fig:Umu4_mN_fa}, we show the $1{}^{}\sigma$ to $3{}^{}\sigma$ contours in the two-parameter plane of $|U_{\mu 4}|^2$ vs $m^{}_N$ with the benchmark choice of other parameters.\,\,Clearly, the neutrino mixing parameter within the range $10^{-10} \lesssim |U_{\mu 4}|^2 \lesssim 10^{-8}$ can account for the latest MiniBooNE results in a broad range of the sterile neutrino mass that is not excluded by the kaon decay experiments.\,\,Notice that there is a small part of the contours in the upper left corner of the figure, where $3\times 10^{-7} \lesssim |U_{\mu 4}|^2 \lesssim 8 \times 10^{-7}$ with $150\,\text{MeV}  \lesssim m^{}_N \lesssim 180\,\text{MeV}$ can fit the data as well.\,\,We also draw the same contours in the $|U_{\mu 4}|^2$ versus the $c^{}_e/f^{}_a$ plane at the bottom panel of Fig.\,\ref{fig:Umu4_mN_fa}, in which $9\times 10^{-3}\,\text{GeV}^{-1} \lesssim c^{}_e/f^{}_a \lesssim 4\times 10^{-2}\,\text{GeV}^{-1}$ with a similar range of $|U_{\mu 4}|^2$ can explain the excess.

In both figures, the upper (lower) portion of the contours corresponds to the short-lived (long-lived) sterile neutrino, where one can approximate the decay probability in Eq.\eqref{PNdec} as
\begin{eqnarray}
P^{}_{N,{\rm dec}}(p^{}_N) 
\,\approx\,
\begin{cases}
\,\displaystyle
\exp\bigg(\hs{-0.08cm}{-}{}^{}L {}^{}\Gamma^{}_N \frac{m^{}_N}{p^{}_N}{}^{}\bigg) 
&\text{for the short-lived sterile neutrino limit} 
\\[0.5cm]
\,\displaystyle
D{}^{}\Gamma^{}_N \frac{m^{}_N}{p^{}_N}
&\text{for the long-lived sterile neutrino limit} 
\end{cases}
~.
\end{eqnarray}
Since $\Gamma^{}_N \propto |U_{\mu 4}|^2$, the number of excess events is increased as the $ |U_{\mu 4}|$ is decreased (increased) for the short-lived (long-lived) sterile neutrino, which explains the behaviors of the contours in these planes.\,\,Accordingly, one can also expect that the lower portion of the contours is very sensitive to $ |U_{\mu 4}|$, and this is because ${\cal S}(Q) \propto |U_{\mu 4}|^2 P^{}_{N,{\rm dec}}(p^{}_N) \propto |U_{\mu 4}|^4$ for the long-lived sterile neutrino.

\section{Discussion and conclusions}\label{sec:5}
First, let us examine the feasibility of our $\ell$ALP model in more detail. In this model, we assume that the $\ell$ALP has a coupling to the sterile neutrino.\,\,Such a coupling can be generated by introducing a complex singlet scalar field $\Phi$ and a pair of left and right chiral fermionic fields $\psi^{}_{L,R}$ of the interactions as~\cite{Alves:2019xpc}
\begin{eqnarray}\label{LPhipsi1}
{\cal L}_{\Phi\psi} \,=\, -\, y^{}_N \big(\Phi {}^{}{}^{} \overline{\psi^{}_L}{}^{} \psi^{}_R + \Phi^\ast {}^{}{}^{} \overline{\psi^{}_R}{}^{} \psi^{}_L\big) ~,
\end{eqnarray}
and those fields are charged under a global axial U$(1)$ symmetry.\,\,After symmetry breaking at the energy scale $\upsilon^{}_\Phi$, where $\upsilon^{}_\Phi$ is the vacuum expectation value of $\Phi$, the angular component of the complex scalar is identified with the $\ell$ALP, $\Phi \supset \upsilon^{}_\Phi + i{}^{}a/\sqrt{2}$.\,\,Thus, Eq.\,\eqref{LPhipsi1}
becomes
\begin{eqnarray}\label{LPhipsi2}
{\cal L}_{\Phi\psi} \,=\, -\, m^{}_N \overline{\psi} {}^{} \psi + \frac{1}{\sqrt{2}}  {}^{} y^{}_N {}^{} a {}^{}\overline{\psi} {}^{} i {}^{} \gamma^5 \psi ~,
\end{eqnarray}
where $m^{}_N = y^{}_N \upsilon^{}_\Phi$.\,\,Comparing Eq.\,\eqref{LPhipsi2} to Eq.\,\eqref{Lint} and identifying $\psi$ with $\nu^{}_\textsf{D}$, we then obtain $y^{}_N = c^{}_N m^{}_N/f^{}_a$.\,\,For our benchmark point, it follows that $y^{}_N \simeq 10^{-3}$ and $\upsilon^{}_\Phi \simeq 250\,\text{GeV}$.

Next, let us discuss the magnitude of the $\ell$ALP-lepton couplings in this model.\,\,For our purpose, we assume that the $\ell$ALP dominantly interacts with electrons.\,\,In other words, the $\ell$ALP couplings to the muon and tau, $c_{\mu,\tau}$, and the $\ell$ALP flavor-changing couplings, say, $c_{e\mu}$, and so forth, are assumed to be negligibly small. Such hierarchy of the axion couplings can be realized in the context of familon/flaxion~\cite{Calibbi:2020jvd,Han:2020dwo,Ema:2016ops}, a pseudo-Nambu-Goldstone boson arising from the spontaneous breaking of a global Froggatt-Nielsen (FN) flavor symmetry, U$(1)^{}_{\rm FN}$~\cite{Froggatt:1978nt}.\,\,For example, one can consider the following effective Yukawa interactions as
\begin{eqnarray}
{\cal L}_{\rm FN} \,=\,
- \, y^{}_{jk} \bigg(\frac{\Phi}{\Lambda}\bigg)^{\hs{-0.13cm}n^{}_{jk}\,} \overline{E^{}_{Lj}} H e^{}_{Rk} + \text{H.c.} ~,
\end{eqnarray}
where $y^{}_{jk}$ is an ${\cal O}(1)$ coefficient, and $\Lambda$ is the cutoff scale of the theory.\,\,Here, $E^{}_{Lj}, H$, and $e^{}_{Rk}$ denote the left-handed lepton doublet, SM Higgs doublet, and right-handed charged lepton singlet, respectively.\,\,The FN charge assignment for those fields is displayed in Table \ref{tab:FN}, from which $n^{}_{jk} = [E^{}_{Lj}]+[e^{}_{Rk}]$.\,\,Then, the breakdown of the FN and electroweak symmetries leads to~\cite{Ema:2016ops}
\begin{eqnarray}
{\cal L}_{\rm FN} \,\supset\,
-\, c^{}_{jk} {}^{}  a {}^{}{}^{} \overline{e^{}_{jL}} {}^{} i {}^{} \gamma^5 e^{}_{kR}
+ \text{h.c.}~,
\quad
c^{}_{jk} \,\equiv\,
y^{}_{jk} {}^{} n^{}_{jk} \frac{\upsilon^{}_{\rm EW}}{\upsilon^{}_\Phi} \bigg(\frac{\upsilon^{}_\Phi}{\Lambda}\bigg)^{\hs{-0.13cm}n^{}_{jk}} ~;
\end{eqnarray}
where we have used $\gamma^5 P_{R, L} = \pm P_{R, L}$, where $\upsilon^{}_{\rm EW} \simeq 174\,\text{GeV}$ is the vacuum expectation value of the SM Higgs field. Therefore, with a proper FN charge assignment, we may achieve the hierarchy of the $\ell$ALP-lepton couplings in our $\ell$ALP model.\,\,However, the construction of a UV completion theory is beyond the scope of this paper, and we leave the detailed study of the model for future work.

In this paper, we have shown that the collimated electron-positron pair produced from the sterile neutrino decay through the $\ell$ALP can account for the recently updated results of the MiniBooNE experiment.\,\,We find that our resulting shapes of the distributions in the neutrino operation mode, especially the angular distribution, are in a good fit with the data. Meanwhile, the total excess event numbers can be explained with the sterile neutrino mass within $150\,\text{MeV}\lesssim m^{}_N \lesssim 380 \,\text{MeV}$ ($150\,\text{MeV}\lesssim m^{}_N \lesssim 180 \,\text{MeV}$) and the neutrino mixing parameter within $10^{-10} \lesssim |U_{\mu 4}|^2 \lesssim 10^{-8}$ ($3\times 10^{-7} \lesssim |U_{\mu 4}|^2 \lesssim 8 \times10^{-7}$).\,\,Moreover, we have checked that our benchmark choice can satisfy constraints from various astrophysical and terrestrial observations.\,\,The scenario could be tested by the searches of the $\ell$ALP from the future colliders and by the sterile neutrino production from the kaon decay facilities.

\renewcommand{\arraystretch}{1.4}
\begin{table}[t!]
\centering 
\begin{tabular}{|c|c|c|c|c|}
\hline
  Field & ~$\overline{E^{}_{Lj}}$~  & ~$H$~ & ~$e^{}_{Rk}$~ & ~$\Phi$~ \\[0.1cm]
\hline
~U$(1)^{}_{\rm FN}$~  & ~$[E^{}_{Lj}]$~  & ~0~  & ~$[e^{}_{Rk}]$~ & ~$-1$~ \\[0.1cm]
\hline
\end{tabular}
\caption{The FN charge assignment of the SM fields and familon, with $j,k =1,2,3$.}
\label{tab:FN}
\end{table}

\section*{Acknowledgments}
S.Y.H. would like to thank P. Ko, J. Kersten, K. Kunio, and C.T. Lu at KIAS for their helpful discussions and useful information.\,\,S.Y.H. and S.Y.T. are grateful to the members of the High Energy Physics Group at National Taiwan Normal University for hospitality, local support, and the beginning of this work.\,\,This work was supported in part by the Ministry of Science and Technology (MOST) of Taiwan under Grant No.\,MOST 109-2112-M-003-004 (C.R.C.), the ORD of National Taiwan Normal University under Grant No.\,109000128 (C.H.V.C.), Korea Institute for Advanced Study under Grant No.\,PG081201 (S.Y.H.), and by JSPS KAKENHI Grant No.\,20J22214 (S.Y.T.).

\section*{Note added}
\begin{figure}[t!]
\vs{-0.2cm}
\centering
\includegraphics[scale=0.483]{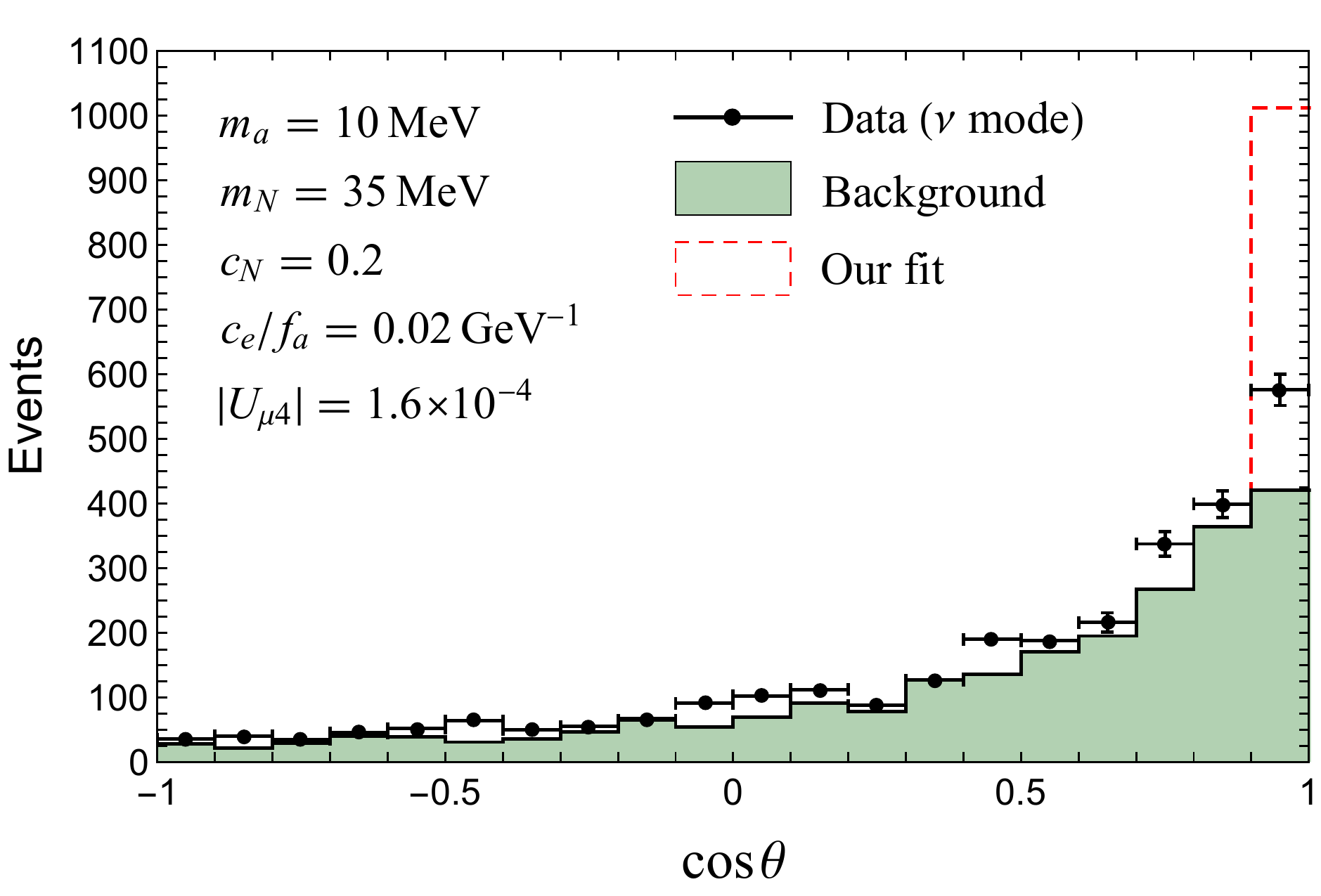}
\\[0.3cm]
\includegraphics[scale=0.48]{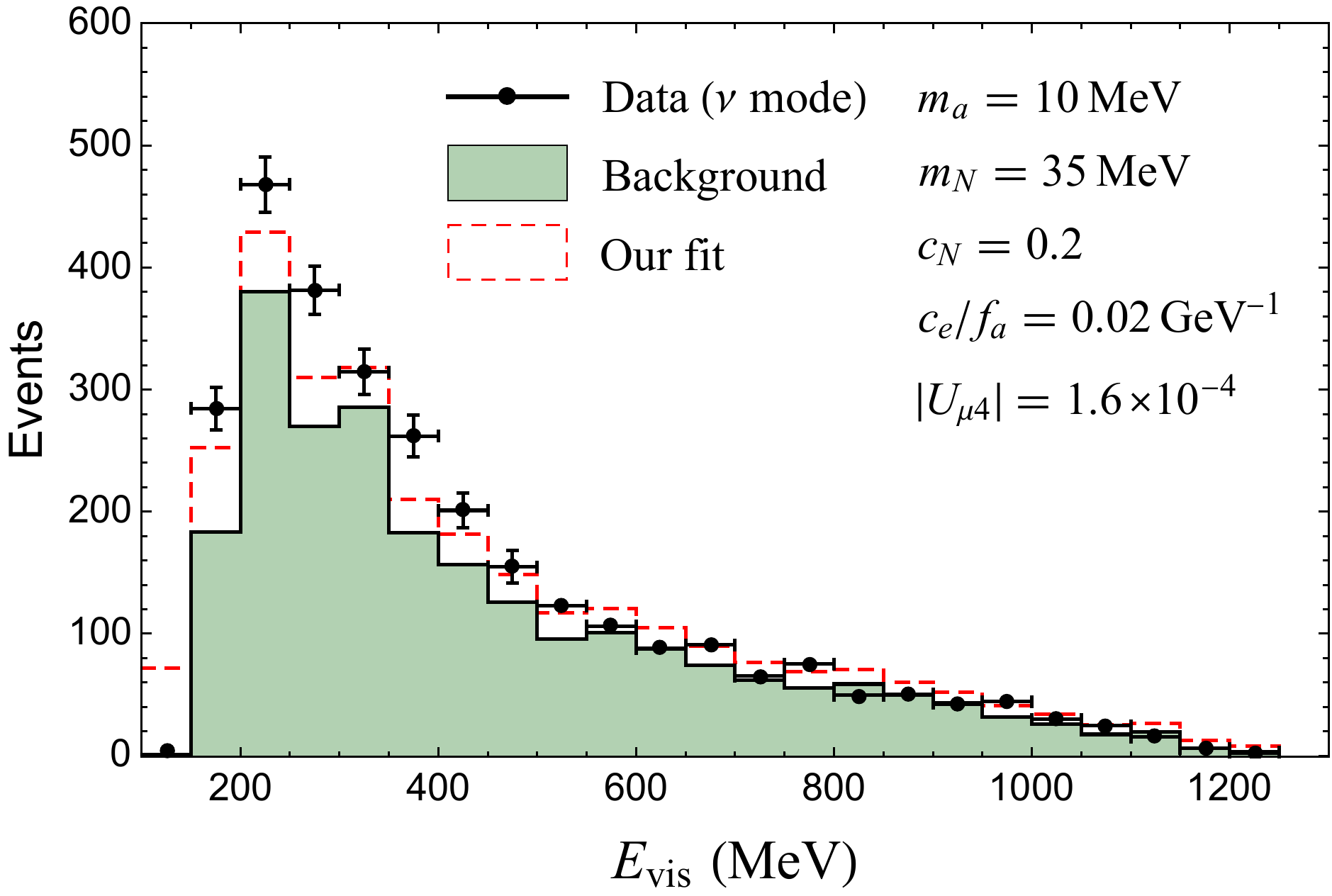}
\caption{The angular and visible energy spectra in the neutrino mode for another benchmark point which takes into account the bunch timing based on the latest MiniBooNE measurement.}
\label{fig:fitting_2}
\end{figure}

We thank  O.\,Fischer for bringing to our attention the fact that most of the excess data event accumulate in the first 8\,ns of the bunch timing based on the latest MiniBooNE measurement\,\cite{MiniBooNE:2020pnu}, which could naively set a mass bound of sterile mass to be less than ${\cal O}$(10\,\text{MeV})\,\cite{Brdar:2020tle}.\,\,Taking the bunch timing into consideration, we find another benchmark point $m^{}_a = 10\,\text{MeV}, m^{}_N = 35\,\text{MeV}, c^{}_N = 0.2, c^{}_e/f^{}_a = 0.02\,\text{GeV}^{-1}, |U_{\mu 4}| = 1.6\times 10^{-4}$, which can satisfy all the constraints in Sec.\,\ref{sec:3} and produce the total excess event number within the $1{}^{}\sigma$ range of the MiniBooNE data.\,\,The angular and visible energy spectra in the neutrino mode for this benchmark point are shown in Fig.\,\ref{fig:fitting_2}, where the shape of the visible energy distribution is similar to the first benchmark point. However, the angular distribution predicted in this case is more forward peaked.

\end{document}